\documentclass[11pt,letter,subeqn,fleqn]{article}
\pdfoutput=1

\usepackage{amsmath, amssymb,amsthm,cite}
\usepackage[normal]{subfigure}
\usepackage{subfig}
\usepackage{graphicx}
\usepackage{float}

\title{Exact Solutions of a Fully Nonlinear Two-Fluid Model}

\author{ 
Alexei F. Cheviakov\footnotemark[1],\vspace{0.5cm}\\
\small\emph{Department of Mathematics and Statistics, University of Saskatchewan, Canada.}\vspace{0.2cm}\\
}

\def\beq{\begin{equation}}
\def\eeq{\end{equation}}
\def\barr{\begin{array}{ll}}
\def\earr{\end{array}}

\def\sg#1{{\rm #1}}
\def\const{\hbox{\rm const}}

\def\grad{\mathop{\hbox{\rm grad}}}

\def\div{{\hbox{\rm div}}}

\def\sn{\mathop{\hbox{\rm sn}}}

\def\vec#1{{\boldsymbol{\rm #1}}} 

\def\cn{\mathop{\hbox{\rm cn}}}
\def\sn{\mathop{\hbox{\rm sn}}}

\newtheorem{theorem}{Theorem}

{\theoremstyle{definition} 

\newtheorem{remark}{Remark}
}

\newcounter{tabnum}

\renewcommand{\div}{\mathrm{div}}

\newcommand{\g}{g}

\begin{document}

\renewcommand{\baselinestretch}{1.1}\small\normalsize

\footnotetext[1]{Alternative English spelling: Alexey Shevyakov. Electronic mail: shevyakov@math.usask.ca}

\maketitle
\numberwithin{equation}{section}

\centerline{\emph{Submitted to the Journal of Fluid Mechanics}}

\begin{abstract}
A nonlinear coupled Choi-Camassa model describing one-dimensional incompressible motion of two non-mixing fluid layers in a horizontal channel has been derived in \cite{choi1999fully}. An equivalence transformation is presented, leading to a special dimensionless form of the system, with the number of constant physical parameters reduced from five to one. A first-order dimensionless ordinary differential equation describing traveling wave solutions is analyzed. Several multi-parameter families of exact closed-form solutions of the Choi-Camassa model are obtained, describing periodic, solitary, and kink-type bidirectional traveling waves.

\end{abstract}


\section{Introduction}

Over the years, multiple simplified models of the systems of Euler and Navier-Stokes fluid dynamics equations have been derived, for specific physical settings, in order to reduce the mathematical complexity of the full set of equations, while retaining essential properties of phenomena of interest and providing sufficient physical insight and computational precision. Basic examples of such simplifications include dimension reductions, linearizations, and more general approximations involving asymptotic relationships. Fundamental nonlinear partial differential equations (PDEs) of mathematical physics, such as Burgers', Korteweg-de Vries (KdV), nonlinear Schr\"{o}dinger, and Kadomtsev-Petviashvili (KP) equations, as well as many other important models like shallow water equations, Camassa-Holm and Degasperis-Procesi equations, arise in the context of fluid dynamics. Importantly, a number of such reduced models exhibit rich mathematical structure, such as integrability,  existence of infinite hierarchies of conservation laws, and solutions in the form of single and/or multiple nonlinear solitary waves (solitons, peakons, etc.). In many cases, exact solutions of reduced models correspond to, and in fact closely describe, physical phenomena. Examples are provided by solitary wave solutions of the KdV equation modeling long waves in shallow channels, periodic solutions of the KP equation modeling crossing swell-type shallow water surface waves, and internal waves. Classical nonlinear wave models are reviewed, for example, in \cite{whitham74,johnson2002camassa, beals2000multipeakons, pelinovsky2007internal}.




In the present work, we consider a nonlinear PDE system, derived by Choi and Camassa in \cite{choi1999fully}, describing nonlinear internal waves in a stratified system of two non-mixing fluids of different densities in a long horizontal channel within the gravity field. The Choi-Camassa (CC) equations have been derived through layer-averaging, under an asymptotic ``shallow water" assumption of a small ratio of  the fluid channel depth to the characteristic wavelength, yet without assuming that wave amplitudes are small compared to the fluid layer depths. The model is one-dimensional, involving four dependent variables (fluid interface displacement, pressure at the interface, and layer-average horizontal velocities) that are functions of time and the spatial coordinate along the channel. The CC equations equations are an extension of the weakly nonlinear model presented in \cite{choi1996weakly}. Further extensions have recently appeared in the literature, including a `regularized' two-fluid model \cite{choi2009regularized}, a mutli-layer model \cite{choi2000modeling}, and a rough bottom model \cite{jo2002dynamics}.

The original paper \cite{choi1999fully} listed several basic local conservation laws of the model, as well as a traveling wave solution ansatz leading to a nonlinear ODE. The latter was shown to admit, for specific parameter relationships, kink- and solitary wave-type traveling wave solutions, which were constructed numerically.  In \cite{camassa2010fullyPer}, periodic traveling wave solutions were obtained as numerical solutions of the same reduced ODE. No exact solutions of the Choi-Camassa model have been reported to date. The main goal of the present paper is the derivation of exact closed-form solutions of the Choi-Camassa model representing traveling solitary waves of depression and elevation, kinks and anti-kinks, and non-harmonic periodic traveling wave displacements of the fluid interface.


The current contribution is organized as follows.

In Section \ref{two-fluid system}, the CC equations of \cite{choi1999fully}  are reviewed, together with important aspects of their derivation, and related results and models. Equivalence transformations are used to recast the four equations, for any set of physical and channel parameters, in a special yet general dimensionless form, which involves only a single dimensionless parameter, namely, the fluid density ratio.

In Section \ref{sec:trw:ODE}, the dimensionless form and the traveling wave ansatz $u(x,t)=u(x-ct)$ is used to derive a first-order nonlinear ordinary differential equation governing all traveling wave-type solutions of the CC model. Importantly, that ODE is independent of the traveling wave speed $c$, and involves a rational polynomial right-hand side.

In Sections \ref{sec:trw:exact:kn:sol} and \ref{sec:trw:exact:denom}, the reduced ODE is utilized to derive families of exact closed-form solutions of the full Choi-Camassa PDE system. These solutions are given by explicit formulae, and include two families of periodic non-harmonic traveling wave solutions, involving elliptic functions. Further, the infinite wavelength limit leads to families of exact expressions for solitary wave and kink/anti-kink (front-type) solutions. Relationships between the solution parameters are investigated, and several examples are considered in detail.

The paper is concluded with Section \ref{sec:end} offering a discussion of properties of the obtained solutions, in particular, in comparison with semi-numerical solutions of similar kinds obtained in the earlier works \cite{choi1999fully, camassa2010fullyPer, gorshkov2011dynamics}, and an overview of related open problems and future research directions.

\section{The Two-Fluid Model, its Properties, and the Dimensionless Form }  \label{two-fluid system}

\subsection{The Governing Equations}

\begin{figure} [htb]
\begin{center}
\includegraphics[width=.65 \textwidth]{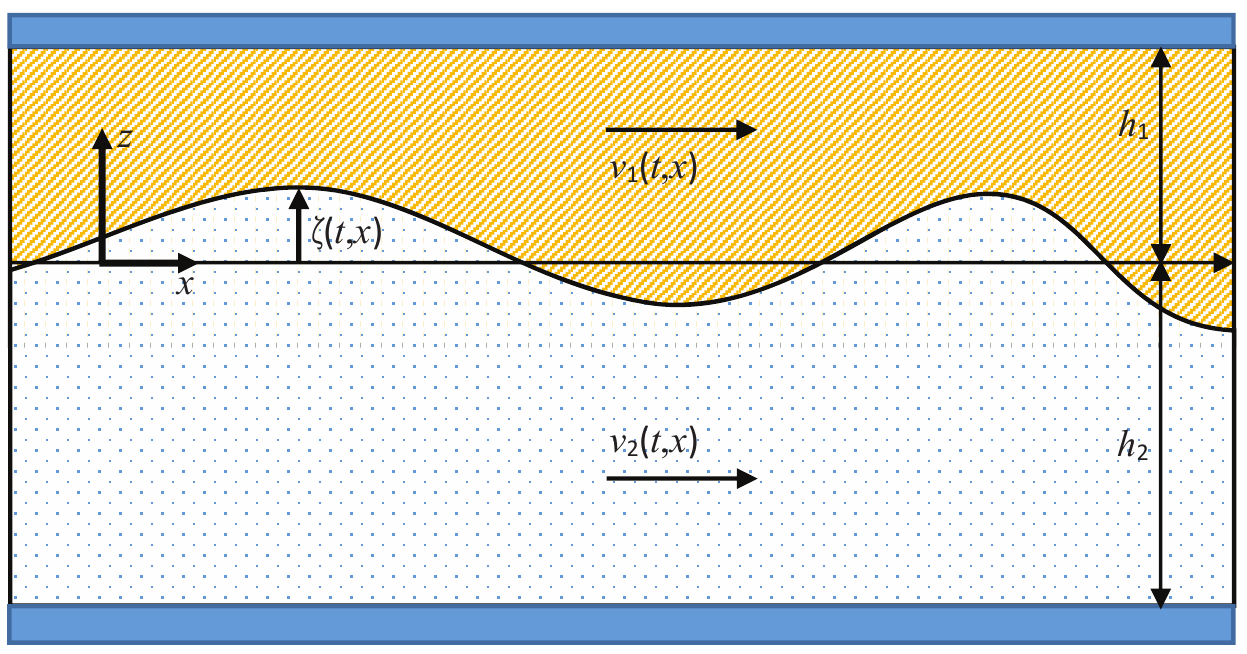}
\caption{The Choi-Camassa model.}
\label{fig01}
\end{center}
\end{figure}

The three-dimensional Euler equations of incompressible inviscid fluid flow of constant density $\rho$ in the gravity field are given by
\begin{equation}\label{NS-Eqn}
\vec{v}_t+(\vec{v}\cdot \nabla) \vec{v} = -  \frac{1}{\rho}\grad p -\vec{g}, \qquad \div\, \vec{v}=0,
\end{equation}
where $\vec{g}=-g\vec{k}$ is the gravitational acceleration, $\vec{v}=(u(t,\vec{x}), v(t,\vec{x}), w(t,\vec{x}))$ is the velocity vector, and $p(t,\vec{x})$ is the fluid
pressure. In \cite{choi1999fully}, Choi and Camassa derived a nonlinear (1+1)-dimensional two-fluid model for an approximate asymptotic description of long waves at the fluid interface.   We briefly overview the relevant notation and elements of the derivation of the Choi-Camassa equations. Consider an irrotational flow within two fluid layers of depths $h_1$, $h_2$ and constant densities $\rho_1$, $\rho_2$ (Figure \ref{fig01}). The condition $\rho_1<\rho_2$ is assumed for a stable stratification.

The incompressible two-dimensional Euler equations in Cartesian coordinates in the $(x,z)$-plane are given by
\begin{equation}\label{eq:2eul:2d}
\barr
u_{x}+w_{z} =0,\\[2ex]
u_{t}+u u_{x}+w u_{z}=-p_{x}/ \rho,\\[2ex]
w_{t}+u w_{x}+w w_{z}=-p_{z}/ \rho -g,
\earr
\end{equation}
with the fluid velocity $\vec{u}=u(t,x,z) \vec{i} + w(t,x,z)\vec{k}$. For the two-fluid model, the PDEs \eqref{eq:2eul:2d} are written for both fluid layers. Denote the flow parameters by $(u,w,p)=(u_i,w_i,p_i)$, $i=1, 2$ for the upper and the lower fluid, respectively. For the two-fluid setup, $z\in [-h_2,h_1]$. In the case of the static equilibrium situation, $0\leq z\leq h_1$ corresponds to the first fluid layer, and $-h_2\leq z\leq 0$ to the second one. No-leak boundary conditions are prescribed at the top and bottom horizontal walls of the channel:
\begin{equation}
w_1(t,x,h_1)= w_2(t,x,-h_2)=0.     \label{KBC}
\end{equation}
Let $\zeta(t,x)$ denote the vertical displacement of the interface between the fluids. The boundary conditions at the interface $z=\zeta(t,x)$ of the two fluids are the continuity of normal velocity and pressure:
\begin{equation}
\zeta_t+u_1\zeta_x=w_1, \quad \zeta_t+u_2\zeta_x=w_2, \quad p_1=p_2.   \label{interface_BC}
\end{equation}
In order to derive the model of interest, an assumption was made that the fluid depth be much smaller than the characteristic length $L$:
\begin{equation}\label{eq:asymp:hL}
h_i/L=\epsilon \ll 1.
\end{equation}
$h_i/L= \epsilon \ll 1$. The continuity equation in \eqref{eq:2eul:2d} yields
\[
w_i/u_i=O(h_i/L)=O(\epsilon) \ll 1.
\]
For finite-amplitude waves, it is assumed that
\begin{equation} \label{eq:fin:amp}
w_i/U_0=O(\zeta/h_i)=O(1),
\end{equation}
where $U_0=(gH)^{1/2}$, $H=h_1+h_2$, is the characteristic speed of the problem.

Denote the actual thicknesses of the fluid layers by
\beq\label{eq:eta_notation}
\eta_1=h_1-\zeta,\qquad \eta_2=h_2+\zeta.
\eeq
As shown in \cite{choi1999fully}, the asymptotic computation for the Euler system \eqref{eq:2eul:2d} with \eqref{interface_BC} leads to a (1+1)-dimensional PDE system for the unknown interface displacement $\zeta(t,x)$, the hydrostatic pressure at the interface $P(t,x)$, and the layer-average (depth-mean) horizontal velocities $v_1(t,x)$, $v_2(t,x)$ of the two fluids defined as
\begin{equation}\label{meanVelocity}
v_1=\frac{1}{\eta_1} \int^{h_1}_\zeta  u_1(t,x,z)  \, dz,   \qquad
v_2=\frac{1}{\eta_2} \int_{-h_2}^\zeta  u_2(t,x,z)  \, dz.
\end{equation}
(We note that in \cite{choi1999fully} and related papers, for the depth-mean velocities, the notation $\bar{u}_i$ was used instead.) Finally, the equations, which we refer to as \emph{the Choi-Camassa (CC) model}, are given by
\begin{subequations}\label{eq:CC}
\beq\label{eq:CC12} 
{\eta_i}_t+(\eta_i v_i)_x=0,
\eeq
\beq\label{eq:CC34}
{v_{i}}_t+v_i{v_{i}}_x+\g\zeta_x= -\dfrac{P_x}{\rho_i} +\dfrac{1}{3\eta_i}\left(\eta_i^3G_i\right)_x + O(\epsilon^4),
\eeq
\end{subequations}
\[
G_i\equiv {v_{i}}_{tx}+v_i{v_{i}}_{xx} -({v_{i}}_x)^2,
\]
\[
i=1,2.
\]
The first-order PDEs \eqref{eq:CC12} are exact; for the purposes of current work, the two remaining PDEs \eqref{eq:CC34} will also be treated as exact, and the  $O(\epsilon^4)$  terms will be omitted.

\subsection{Discussion of the Choi-Camassa model \eqref{eq:CC}}


\medskip\noindent \emph{1) Generalizations and related models.}  First, it is worth mentioning that the CC equations \eqref{eq:CC} were derived under the scaling assumption \eqref{eq:asymp:hL}, not assuming, for example, that the wave amplitudes are small compared to the channel depth. If the latter assumption is imposed, the PDEs \eqref{eq:CC} lead to the Boussinesq approximation, and further, the KdV equation for the case of unidirectional waves \cite{choi1999fully}.

The PDE system \eqref{eq:CC} is in fact an extension the earlier weakly nonlinear model presented in \cite{choi1996weakly}. It was further generalized in \cite{choi2000modeling} where a closed channel with $N\geq 2$ fluid layers was considered. A `regularized' version of the CC model where instead of layer-mean horizontal velocities, velocity values at constant $z$ have been used, has been suggested in \cite{choi2009regularized}. A modification of the CC system to account for an uneven bottom topography has been suggested in \cite{jo2002dynamics}.

\medskip\noindent \emph{2) Asymptotic horizontal velocity estimates.} The velocity components of a solution of the Choi-Camassa equations are the layer-averaged velocity values \eqref{meanVelocity}. Approximate values of actual velocities at a fixed value of $z$ can be readily derived. For irrotational flows, the horizontal velocity components can be written as \cite{whitham74}
\beq\label{eq:u:Whitham}
u_i (t, x, z) = u^{(0)}_i -\tfrac{1}{2}(z\mp h_i)^2 u^{(0)}_{i\,xx} + O(\epsilon^4),\qquad i=1,2,
\eeq
where $u^{(0)}_1(t, x)$, $u^{(0)}_2(t, x)$ are the horizontal velocity values at the top and bottom channel boundaries $z=h_1$, $z=-h_2$, respectively, and the second term is $O(\epsilon^2)$ \cite{choi1999fully, choi2009regularized}.
Further, as pointed out in \cite{choi2009regularized}, the layer-averaged velocities can be computed from \eqref{meanVelocity}, \eqref{eq:u:Whitham} to yield
\beq\label{eq:umean:Jo}
v_i  = u^{(0)}_i -\tfrac{1}{6}\eta_i^2 u^{(0)}_{i\,xx} + O(\epsilon^4).
\eeq
It is straightforward to show that to the same precision, a backward relationship holds (cf. \cite{whitham74}):
\beq\label{eq:umean:AFS}
u^{(0)}_i  = v_i  + \tfrac{1}{6}\eta_i^2 v_{i\,xx} + O(\epsilon^4).
\eeq
One consequently has an asymptotic expression
\beq\label{eq:ui:thru:vCC}
u_i (t, x, z) = v_i  + \left(\tfrac{1}{6}\eta_i^2  -\tfrac{1}{2}(z\mp h_i)^2\, \right)v_{i\,xx} + O(\epsilon^4)
\eeq
for the horizontal velocities $u_i(t, x, z)$ within the channel in terms of the mean velocity of the corresponding fluid layer.

\medskip\noindent \emph{3) Boundary conditions in an infinite channel.} A physically natural initial value problem for the Choi-Camassa system \eqref{eq:CC} would be, for example, one stated for $x\in \mathbb{R}$, with appropriate initial conditions, and boundary conditions at infinity:
\beq\label{eq:CC:BCs:inf}
\barr
v_1(t,x)\to V^{-}_1,\quad  v_2(t,x)\to V_2^{-}, \quad \zeta(t,x)\to \zeta^{-}~~\text{as}~~x\to-\infty,\\
v_1(t,x)\to V^{+}_1,\quad  v_2(t,x)\to V_2^{+}, \quad \zeta(t,x)\to \zeta^{+}~~\text{as}~~x\to+\infty.
\earr
\eeq
For a kink-type solution, for example, one would have $\zeta^{-}\ne\zeta^{+}$, whereas for a solitary traveling wave or a multi-soliton situation, $\zeta^{-}=\zeta^{+}$. Alternatively, for example, for the case of periodic traveling wave solutions considered below, it may be appropriate to consider PDEs \eqref{eq:CC} in a finite interval $I\subset \mathbb{R}$ with periodic boundary conditions.

As noted in \cite{choi1999fully}, the exclusion of $\zeta_t$ from the first two PDEs \eqref{eq:CC12} leads to the formula
\[
\dfrac{\partial}{\partial x}(\eta_1 v_1+\eta_2 v_2)=0,
\]
which, under zero boundary conditions at infinity, yields
\beq\label{eq:notused}
\dfrac{v_2}{v_1}=-\dfrac{\eta_1}{\eta_2}.
\eeq
However, in the current manuscript, we do not make any \emph{a priori} assumptions about boundary conditions; in particular, the relationship \eqref{eq:notused} is not used.

\medskip\noindent \emph{4) Symmetry properties and traveling wave solutions.} It is evident that the symmetry group of the PDE system \eqref{eq:CC} includes translations in $x$ and $t$, translation of the pressure by an arbitrary function of time, and the Galilei group:
\beq\label{eq:symmsCC}
\barr
x^*=x + x_0+ Ct,\qquad t^*=t + t_0,\qquad (v_{i})^*=v_{i} + C,\qquad P^*=P + P_0(t),\\[1.5ex]
x_0, t_0, C = \const.
\earr
\eeq
The Galilei transformation can be used, for example, to set $V^{-}_1=0$ in the boundary conditions \eqref{eq:CC:BCs:inf}, without loss of generality.

The space-time translation symmetry of the Choi-Camassa system leads to the existence of the traveling wave solution ansatz, which is considered in detail in Section \ref{sec:trw:ODE} below, and used for the construction of several families of exact solutions of the CC model in the following Sections \ref{sec:trw:exact:kn:sol} and \ref{sec:trw:exact:denom}.

An important property related with the symmetry structure of the Choi-Camassa model \eqref{eq:CC}, considered in Section \ref{sec:dimless}, consists in an existence of equivalence transformations that lead to a dimensionless form and the reduction of the number of parameters.

\subsection{The Dimensionless Form and Parameter Reduction}\label{sec:dimless}

The Choi-Camassa system has been originally derived using dimensionless variables for the purpose of asymptotic comparisons, but presented in the dimensional form \eqref{eq:CC} containing five constant physical parameters
\beq\label{eq:CC:PhysParam}
g,~\rho_1,~\rho_2,~h_1, ~h_2.
\eeq
We now derive a different dimensionless form of the PDE system \eqref{eq:CC}, with a goal of minimization of the number of physical constants in the system. For this purpose, it is convenient to use the total channel depth
$H=h_1+h_2$
as the length parameter, and the quantity
\beq\label{eq:CC:nondim:Z}
\hat{Z}=\dfrac{h_1-\zeta}{H} \equiv \dfrac{\eta_1}{H},\qquad 0<\hat{Z}<1,
\eeq
as a dependent variable instead of $\zeta$. The physical meaning of $\hat{Z}$ is the relative depth of the top fluid level. Define the further dimensionless variables $\hat{x}$, $\hat{t}$, $\hat{v}_1$, $\hat{v}_2$, $\hat{P}$ by formulas
\beq\label{eq:CC:nondim:transf}
\barr
t=Q_t \,\hat{t},\qquad x=Q_h \,\hat{x},\\[2ex]
P(t,x)=Q_P \hat{P}(\hat{t},\hat{x}),\qquad v_i(t,x)=Q_i \hat{v}_i(\hat{t},\hat{x}),\quad i=1,2.
\earr
\eeq
The respective constants $Q$ can be chosen so that the governing equations in terms of variables with hats will only involve \emph{a single parameter}
\beq\label{eq:S:dens:ratio}
S=\dfrac{\rho_1}{\rho_2},\qquad 0<S<1.
\eeq
The scaling constants are given by
\beq\label{eq:CC:nondim:const}
Q_h=H, \quad Q_t=\sqrt{\dfrac{H}{g}},\quad Q_1=Q_2=\sqrt{{g}{H}}, \quad Q_P=\rho_1 g H,
\eeq
and lead to the \emph{dimensionless Choi-Camassa system}, given by
\begin{subequations}\label{eq:CC:nondim:sys}
\begin{equation}
\hat{Z}_{\hat{t}}+(\hat{Z} \hat{v}_{1})_{\hat{x}}=0,\label{eq:CC:nondim:sys:1}
\end{equation}
\begin{equation}\label{eq:CC:nondim:sys:2}
\hat{Z}_{\hat{t}}+(\hat{Z}\hat{v}_{2})_{\hat{x}} - ({\hat{v}_2})_{\hat{x}}=0,
\end{equation}
\begin{equation}\label{eq:CC:nondim:sys:3}
\hat{v}_{1{\hat{t}}}+\hat{v}_1\hat{v}_{1\,{\hat{x}}}-\hat{Z}_{\hat{x}}+ \hat{P}_{\hat{x}} - \hat{Z} \hat{Z}_{\hat{x}} \hat{G}_1 -\tfrac{1}{3}\hat{Z}^2 \hat{G}_{1\,{\hat{x}}} =0,
\end{equation}
\begin{equation}\label{eq:CC:nondim:sys:4}
\hat{v}_{2{\hat{t}}}+\hat{v}_2\hat{v}_{2\,{\hat{x}}}-\hat{Z}_{\hat{x}}+S \hat{P}_{\hat{x}} -\tfrac{1}{3}(1-\hat{Z})^2 \hat{G}_{2\,{\hat{x}}}+(1-\hat{Z})\hat{Z}_{\hat{x}} \hat{G}_2 =0,
\end{equation}
\[
\hat{G}_i\equiv \hat{v}_{i\,tx}+\hat{v}_i \hat{v}_{i\,xx} -(\hat{v}_{i\,\hat{x}})^2,\qquad i=1,2.
\]
\end{subequations}
The dimensionless form \eqref{eq:CC:nondim:sys} is preferable to the original Choi-Camassa equations \eqref{eq:CC}, for example, in analyses involving classifications, such as symmetry and conservation law classifications, and stability analysis \cite{BCABook}.

The `price' paid for the reduction of the number of parameters is the loss of the apparent likeness between the pairs of equations \eqref{eq:CC} for each fluid layer. This similarity is, however, not perfect, in particular, due to the difference of signs in \eqref{eq:CC34}. As a result, there is no `fluid interchange' equivalence transformation that would exchange, for example, $(v_1,\rho_1) \leftrightarrow (v_2,\rho_2)$, $\zeta \leftrightarrow -\zeta$, etc.

Since the Choi-Camassa system \eqref{eq:CC} is mapped into the dimensionless form \eqref{eq:CC:nondim:sys} for any set of physical parameters \eqref{eq:CC:PhysParam}, it follows that for the original system \eqref{eq:CC}, there exist \emph{equivalence transformations} \cite{ovsiannikov2014group, BCABook} that freely modify the parameters \eqref{eq:CC:PhysParam}, while preserving the density ratio \eqref{eq:S:dens:ratio}.

\section{The Traveling Wave Ansatz}\label{sec:trw:ODE}

We start with a brief derivation of an ordinary differential equation describing bidirectional constant-speed traveling waves for the dimensionless Choi-Camassa system  \eqref{eq:CC:nondim:sys}. The existence of this important ansatz follows from the invariance of the PDEs \eqref{eq:CC:nondim:sys} under the point transformations \eqref{eq:symmsCC},
and consequently, under a combined point symmetry with the generator
\[
\sg{X}={\hat{c}}\dfrac{\partial}{\partial {\hat{x}}} + \dfrac{\partial}{\partial {\hat{t}}},
\]
for an arbitrary dimensionless wave speed ${\hat{c}}=\const$. The invariants of $\sg{X}$ are all dependent variables of the problem, and the traveling wave coordinate
\beq\label{eq:r_trav}
\hat{r}=\hat{r}(t,x)={\hat{x}}-\hat{c}\hat{t}+{\hat{x}}_0 = \dfrac{1}{H}(x-c t + x_0),
\eeq
${\hat{x}}_0=\const$, $x_0=H\hat{x}_0$, $c=\hat{c}\sqrt{gH}$.
Traveling wave solutions of \eqref{eq:CC:nondim:sys} are consequently sought in the form $\hat{Z}({\hat{t},{\hat{x}}})=\hat{Z}(\hat{r})$, etc.,
leading to a system of four ODEs. This system can be reduced to a single first-order ODE; our derivation proceeds somewhat differently from the one in \cite{choi1999fully}.

Within the current section, we use primes to denote the derivatives of the dependent variables $\hat{Z}$, $\hat{v}_1$, $\hat{v}_2$, $\hat{P}$ with respect to $\hat{r}$. The substitution of the traveling wave ansatz  into the first two ODEs of \eqref{eq:CC:nondim:sys} yields
\beq\label{eq:trw:ode1:2}
{\hat{c}} \hat{Z}'=(\hat{Z}\hat{v}_1)' = (\hat{Z}\hat{v}_2)'  - \hat{v}_2'.
\eeq
The general solution for the average velocity expressions is given by
\beq\label{eq:trw:v1v2}
\hat{v}_1={\hat{c}}+\dfrac{C_1}{\hat{Z}}, \qquad \hat{v}_2={\hat{c}} +\dfrac{C_2}{1-\hat{Z}},
\eeq
where $C_1,C_2$ are arbitrary constants. The dimensionless velocity expressions \eqref{eq:trw:v1v2} are regular functions, since physically, $0<\hat{Z}<1$. Substituting \eqref{eq:trw:v1v2} into the ODE version of \eqref{eq:CC:nondim:sys:3}, one explicitly finds the pressure in terms of $\hat{Z}$:
\beq\label{eq:trw:P}
\hat{P}=\hat{P_0} + \hat{Z} -\dfrac{C_1^2}{6\hat{Z}^2}\Big(2 \hat{Z} \hat{Z}'' -(\hat{Z}')^2+3\Big),\qquad \hat{P_0}=\const.
\eeq
Using \eqref{eq:trw:v1v2}, \eqref{eq:trw:P} in the final ODE following from \eqref{eq:CC:nondim:sys:4} yields a rather complicated third-order ODE for $\hat{Z}(\hat{r})$, which we denote by $E_4[\hat{Z}]$. To reduce its order, we seek conservation law multipliers (integrating factors) of this equation in the form $\Lambda=\Lambda(\hat{r},\hat{Z})$,
through the direct construction method (see, e.g., \cite{BCABook, cheviakov2007gem, cheviakov2010computation, kallendorf2012conservation, kelbin2013new, cheviakov2014generalized}). Two integrating factors are immediately found, given by
\[
\Lambda_1=\hat{Z}^{-3}(1-\hat{Z})^{-3},\qquad \Lambda_2=\hat{Z}^{-2}(1-\hat{Z})^{-3}.
\]
The respective linearly independent first integrals satisfy
\[
\Lambda_i E_4 = \dfrac{d}{dr}\Phi_i[\hat{Z}],\qquad i=1,2,
\]
and are given by
\beq\label{eq:trw:Phi1}
\barr
\Phi_1[\hat{Z}] &=  -\dfrac{1}{2\hat{Z}^2(1-\hat{Z})^2}\Big[ 2\hat{Z}(1-\hat{Z})(\alpha_1 \hat{Z}+\alpha_0 )\hat{Z}'' \\[2ex]
&\qquad + \Big( \alpha_0(1-2\hat{Z}) - \alpha_1 \hat{Z}^2\Big)\Big(3-(\hat{Z}')^2\Big) + 6(1-S)\hat{Z}^3(1-\hat{Z})^2 \Big] \\[2ex]
&= K_1 = \const,
\earr
\eeq
\beq\label{eq:trw:Phi2}
\barr
\Phi_2[\hat{Z}] &=  -\dfrac{1}{2\hat{Z} (1-\hat{Z})^2}\Big[ 2\hat{Z}(1-\hat{Z})(\alpha_1 \hat{Z}+\alpha_0 )\hat{Z}'' \\[2ex]
&\qquad + \Big( \alpha_1  \hat{Z} (1-2\hat{Z}) +\alpha_0(2-3\hat{Z})\Big) \Big(3-(\hat{Z}')^2\Big) + 3(1-S)\hat{Z}^3(1-\hat{Z})^2 \Big] \\[2ex]
&= K_2 = \const,
\earr
\eeq
where the short-hand notation for constant combinations
\beq\label{eq:C12alpha12}
\alpha_0=C_1^2 S,\qquad \alpha_1  = C_2^2-\alpha_0
\eeq
has been used. In \eqref{eq:trw:Phi1}, \eqref{eq:trw:Phi2}, $K_1$ and $K_2$ are arbitrary constants corresponding to the choice of the boundary conditions of the original third-order ODE $E_4[\hat{Z}]$. The two first integrals \eqref{eq:trw:Phi1}, \eqref{eq:trw:Phi2} are now used to reduce the order of the ODE at hand by two. This can be done by substitution of $\hat{Z}''$ from one expression into the other, or by noticing that the linear combination $\Phi_1[\hat{Z}]\hat{Z}-\Phi_2[\hat{Z}]$ does not involve $\hat{Z}''$. One arrives at the first-order ordinary differential equation
\beq\label{eq:ODE:Z2}
({\hat{Z}}')^2=\dfrac{A_4 \hat{Z}^4 + A_3 \hat{Z}^3 + A_2 \hat{Z}^2 + A_1 \hat{Z}  + A_0}{\alpha_1 \hat{Z} +\alpha_0},
\eeq
where
\beq\label{eq:ODE:Z2Ai}
\barr
A_4=3(1-S), \quad A_3=2K_1-A_4, \\[2ex]
A_2=-2(K_1+K_2), \quad A_1=2K_2+3\alpha_1, \quad A_0=3\alpha_0.
\earr
\eeq
Overall, the family of ODEs \eqref{eq:ODE:Z2} involves four independent constant parameters. For example, one may choose
\[
\alpha_0\geq 0, \quad\alpha_1, A_2, A_3\in \mathbb{R}
\]
as arbitrary constants. Then one has
\beq\label{eq:ODE:Z2Ai:A1}
A_1=3\alpha_1-(A_2+A_3+A_4),
\eeq
and the only additional restriction following from \eqref{eq:C12alpha12} is given by $\alpha_0+\alpha_1\geq 0$. The coefficient $A_4>0$ is physically defined by the fluid density ratio $S$ according to \eqref{eq:S:dens:ratio}. Further, for a nontrivial flow in which the average velocities \eqref{eq:trw:v1v2} are not constant at the same time, one requires $C_1C_2\ne 0$, hence the denominator of \eqref{eq:ODE:Z2} does not vanish.

The physical solution additionally depends on the arbitrary constant parameter ${\hat{c}}$. In terms of $\alpha_0$, $\alpha_1$, ${\hat{c}}$, $S$, the dimensionless velocities are given by
\beq\label{eq:trw:v1v2:alpha}
\hat{v}_1={\hat{c}}\pm\dfrac{\sqrt{\alpha_0/S}}{\hat{Z}}, \qquad \hat{v}_2=\hat{c} \pm  \dfrac{\sqrt{\alpha_0+\alpha_1}}{1-\hat{Z}},
\eeq
where the signs $\pm$ can be independently chosen in both velocity expressions to yield independent solutions of the Choi-Camassa model. Indeed, it is straightforward to verify that the model equations \eqref{eq:CC} are satisfied for any velocity expressions arising from \eqref{eq:trw:v1v2} with $C_1$, $C_2$ satisfying \eqref{eq:C12alpha12}.

\medskip
The nonlinear autonomous ODEs \eqref{eq:ODE:Z2} with a rational right-hand side do not belong to any well-studied DE class, and a closed form of their general solution is not known. An implicit general solution of \eqref{eq:ODE:Z2} is readily written but is not practically useful. (We note that if there is a common root for the numerator and denominator of the right-hand side of \eqref{eq:ODE:Z2}, the resulting ODEs with a cubic right-hand side are in the class of Jacobi-like equations; e.g., \cite{whitham74}.) Since the ODE is first-order, it admits infinite sets of point symmetries and integrating factors. To find the latter, however, is generally a more difficult problem than to solve the ODE itself (e.g., \cite{BCABook}).

A dimensional ODE analogous to \eqref{eq:ODE:Z2} has been obtained in \cite{choi1999fully}, and earlier, also in the context of internal waves, in \cite{miyata1985internal}.  The zeroes of the numerator were analyzed in \cite{choi1999fully} using a diagram which established parameter ranges and appropriate initial conditions to numerically produce solitary wave-type solutions, which were argued to compare well with experimental data of \cite{koop1981investigation}. In \cite{camassa2010fullyPer, gorshkov2011dynamics}, further analysis led to finding appropriate conditions for numerical kink and periodic traveling wave solutions. Numerical comparisons of \cite{camassa2010fullyPer} between the solutions of the full Euler equations in a two-fluid channel, the CC system, and the KdV model demonstrated a reasonable agreement between the first two models, and a relatively poor approximation provided by the latter.

\section{Cnoidal and Solitary Waves}\label{sec:trw:exact:kn:sol}

Solitary and cnoidal waves are known to exist for the Korteweg-de Vries equation and a number of other nonlinear models. We show that physical solutions of such kinds also arise from the Camassa-Choi traveling wave ODE \eqref{eq:ODE:Z2}. The following theorem holds.

\begin{theorem} \label{th:sol:snSQ}
The family of ODEs \eqref{eq:ODE:Z2} admits exact solutions in the form
\beq\label{eq:th:sol:elliptic:Z:snSQ}
\hat{Z}(\hat{r})={B_{1}}\,{\sn}^2(\gamma \,\hat{r}, k)+B_{2},
\eeq
for arbitrary constants $k, B_1, B_2$. The remaining constants $\gamma$ and $\alpha_{1,2}$ are given by one of the two relationships \eqref{eq:th:snSQ:flas1}, \eqref{eq:th:snSQ:flas2} listed in  Appendix \ref{app:Th:snSQ}.
\end{theorem}
Theorem \ref{th:sol:snSQ} is proven by a direct substitution. One consequently has \emph{two families} of exact solutions of the dimensionless Choi-Camassa system \eqref{eq:CC:nondim:sys}, each depending on the arbitrary constants $k, B_1, B_2, \hat{c}$, as well as arbitrarily prescribed physical parameters $S=\rho_1/\rho_2, h_1, h_2, g$. The dimensional solutions of the Choi-Camassa system \eqref{eq:CC} are computed as follows. First, as per the definition of $\hat{Z}$ in \eqref{eq:CC:nondim:Z} and the formula \eqref{eq:th:sol:elliptic:Z:snSQ}, one has
\beq\label{eq:trw:snSQ:sol:ZetaNonspec}
\zeta (x,t)= h_1-H \hat{Z} = (h_1-HB_2) - HB_1\,{\sn}^2(\gamma \,\hat{r}(x,t), k).
\eeq
For both solution families, the pressure, according to \eqref{eq:CC:nondim:transf}, \eqref{eq:CC:nondim:const}, \eqref{eq:trw:P}, is given by
\beq\label{eq:trw:snSQ:sol:P}
P(x,t)=\rho_1 g H \hat{P}(\hat{r}(x,t)).
\eeq
The dimensional average velocities have different expressions for the two cases that arise.

\medskip\noindent\textbf{Case 1.} For the relationship \eqref{eq:th:snSQ:flas1} between the solution parameters, one has $\alpha_0+\alpha_1=C_2^2=0$. The mean-layer velocities computed from \eqref{eq:trw:v1v2} and \eqref{eq:CC:nondim:transf} are given by
\beq\label{eq:trw:v1v2:solA}
{v}_1(x,t)=\sqrt{{g}{H}}\left({\hat{c}}\pm \dfrac{\sqrt{\alpha_0/S}}{B_1\,{\sn}^2(\gamma \,\hat{r}(x,t), k)+B_2}\right), \qquad {v}_2(x,t)={\hat{c}}\sqrt{{g}{H}} = \const,
\eeq
where different choices of the sign yield different admissible forms of ${v}_1(x,t)$.

\medskip\noindent\textbf{Case 2.}
For the second parameter relationship \eqref{eq:th:snSQ:flas2}, one has $C_1=\alpha_0=0$, $C_2=\pm\sqrt{\alpha_1}$, the solution has a constant upper fluid layer-average velocity ${v}_1(x,t)$ and two possible expressions for ${v}_2(x,t)$:
\beq\label{eq:trw:v1v2:solB}
{v}_1(x,t)={\hat{c}}\sqrt{{g}{H}} = \const,\qquad
{v}_2(x,t)=\sqrt{{g}{H}}\; \left(\hat{c}\pm \dfrac{\sqrt{\alpha_1}}{1-B_1\,{\sn}^2(\gamma \,\hat{r}(x,t), k)-B_2}\right).
\eeq

\begin{remark}
The constant value of a corresponding layer-average velocity $v_i=\const$ in Cases 1 and 2 above implies, through the asymptotic expression \eqref{eq:ui:thru:vCC}, that the corresponding horizontal fluid velocity component $u_i(t,x,z)=O(\epsilon^4)$. We note that in the papers \cite{choi1999fully, camassa2010fullyPer} where solitary wave profiles were obtained numerically, the behaviour of the layer-average velocities was not discussed.
\end{remark}

\begin{remark}
In order to describe elevation or depression wave trains positioned directly above or below the fluid interface $\zeta=0$, one can choose
\begin{subequations}\label{eq:trw:snSQ:sol:Zeta:CN2}
\beq\label{eq:trw:snSQ:sol:Zeta:cn2:B2}
B_2=\dfrac{h_1}{H}-B_1.
\eeq
This leads to the interface displacement formula
\beq\label{eq:trw:snSQ:sol:Zeta}
\zeta(x,t)= HB_1\,{\cn}^2(\gamma \,\hat{r}(x,t), k).
\eeq
\end{subequations}
\end{remark}

\subsection{Periodic cnoidal waves} \label{sec:trw:exact:kn:kn}

\begin{figure}[ht]
\begin{center}\hspace*{-2cm}
\includegraphics[height = 5cm,clip]{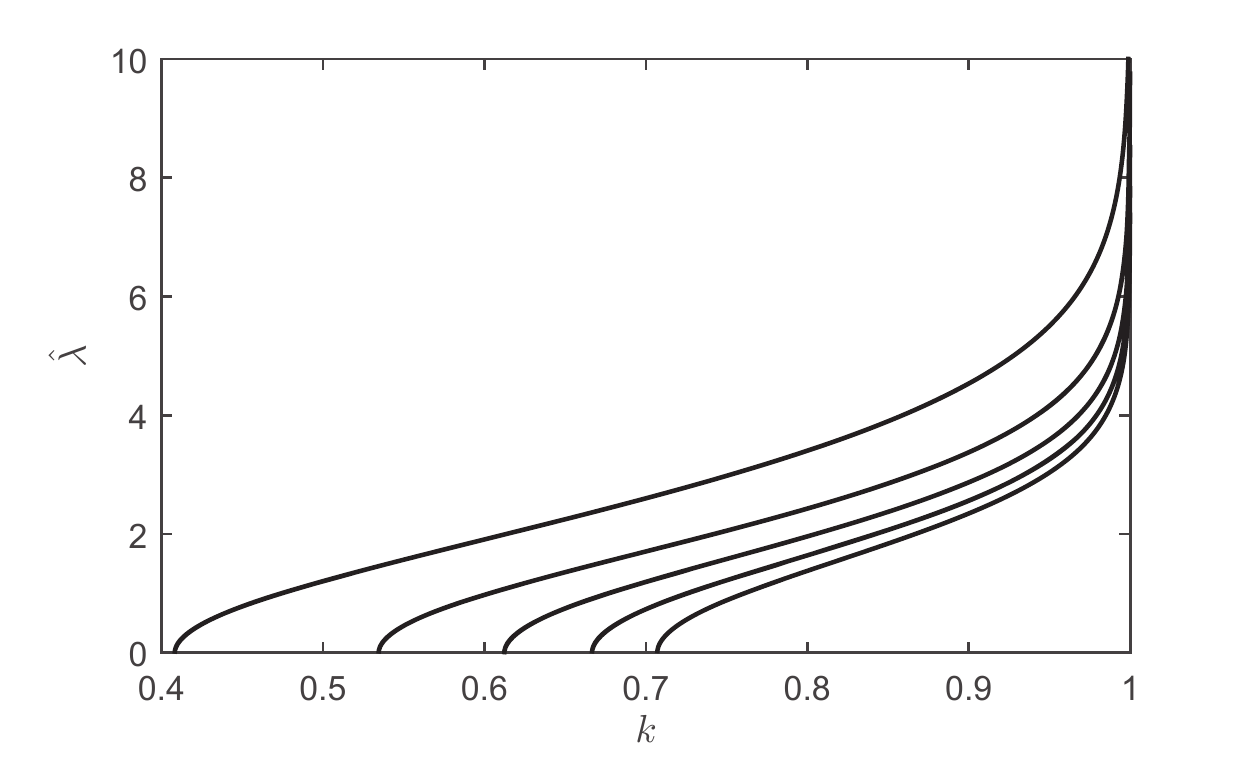}
\caption{Dependence of the dimensionless wavelength $\hat{\lambda}$ of the special cnoidal wave \eqref{eq:trw:snSQ:sol:Zeta:CN2} on the elliptic function parameter $k$, for $h_1/H=2/5$, for Case 1 (formulas \eqref{eq:th:snSQ:flas1}). From top to bottom: curves for $B_1=-0.4, -0.32, -0.24, -0.16, -0.08$. }
\label{fig:sol:sn2per:period}
\end{center}
\end{figure}

Periodic cnoidal-type solutions of the Choi-Camassa system \eqref{eq:CC}  arise from the formula \eqref{eq:th:sol:elliptic:Z:snSQ} for $0<k<1$. The period of the elliptic sine $\mbox{sn}(x,k)$ is given by
\beq\label{eq:sn:T}
\tau=\dfrac{2\pi}{M(1,\sqrt{1-k^2})},
\eeq
where $M(a,b)$ denotes the Gauss' algebraic-geometric mean of $a,b$; for the function $\mbox{sn}^2(x,k)$, the period equals $\tau/2$. The dimensionless and the dimensional wavelength of the exact solutions arising from \eqref{eq:th:sol:elliptic:Z:snSQ} are consequently given by
\beq\label{eq:th:sol:elliptic:Z:snSQ:Wavelength}
\hat{\lambda}=\dfrac{\pi}{\gamma\,M(1,\sqrt{1-k^2})},\qquad \lambda=H \hat{\lambda}.
\eeq
For the cnoidal wave solutions \eqref{eq:th:sol:elliptic:Z:snSQ}, $\gamma=\gamma(B_1,B_2,k)$ according to \eqref{eq:th:snSQ:flas1} or \eqref{eq:th:snSQ:flas2}. In particular, $\gamma$ is independent of the fluid densities since it does not involve the density ratio $S$. Sample plots of $\gamma(B_1,k)$ in Case 1 (formulas \eqref{eq:th:snSQ:flas1}), for the choice of $B_2$ according to \eqref{eq:trw:snSQ:sol:Zeta:cn2:B2}, with $h_1/H=2/5$, are given in Figure \ref{fig:sol:sn2per:period}. The limit $k\to 1^-$, $\hat{\lambda}\to +\infty$ corresponds to the cnoidal-solitary wave transition.

%

\begin{figure}[ht]
\begin{center}
\subfigure[~]{\includegraphics[width = 7cm,clip]{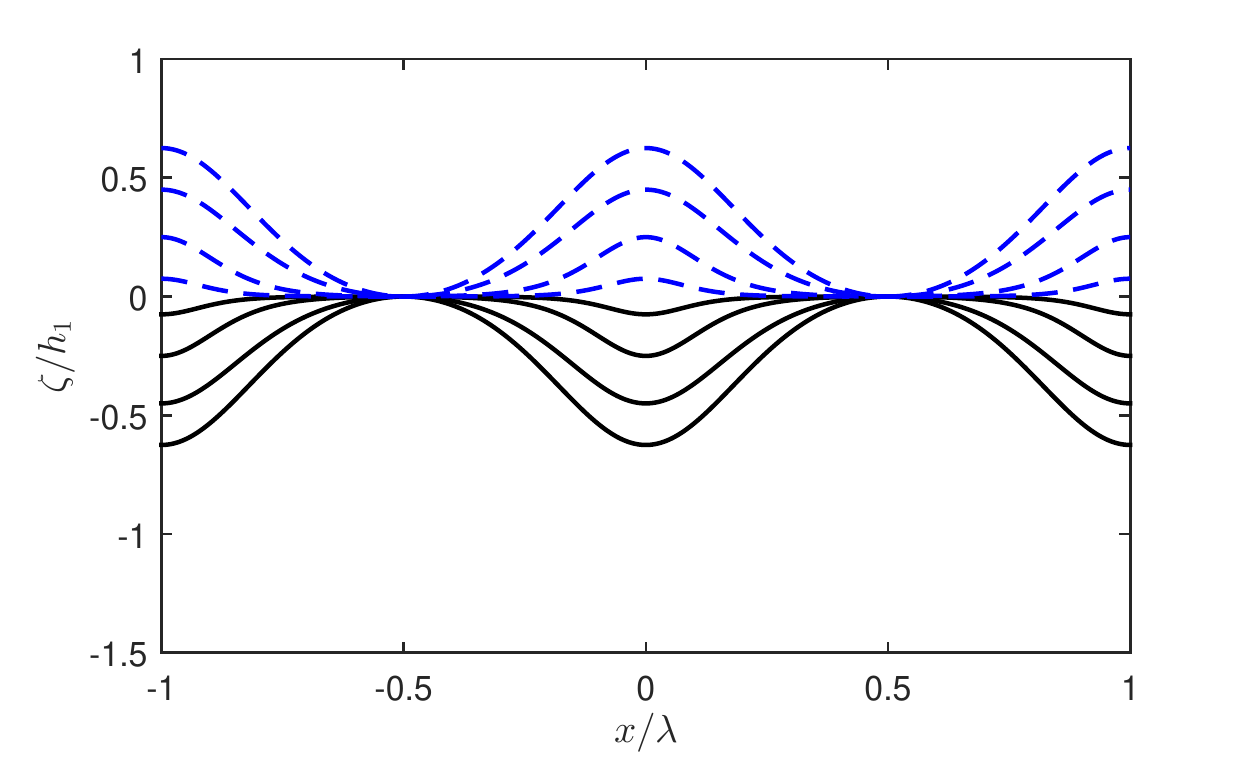} } \subfigure[~]{\includegraphics[width = 7cm,clip]{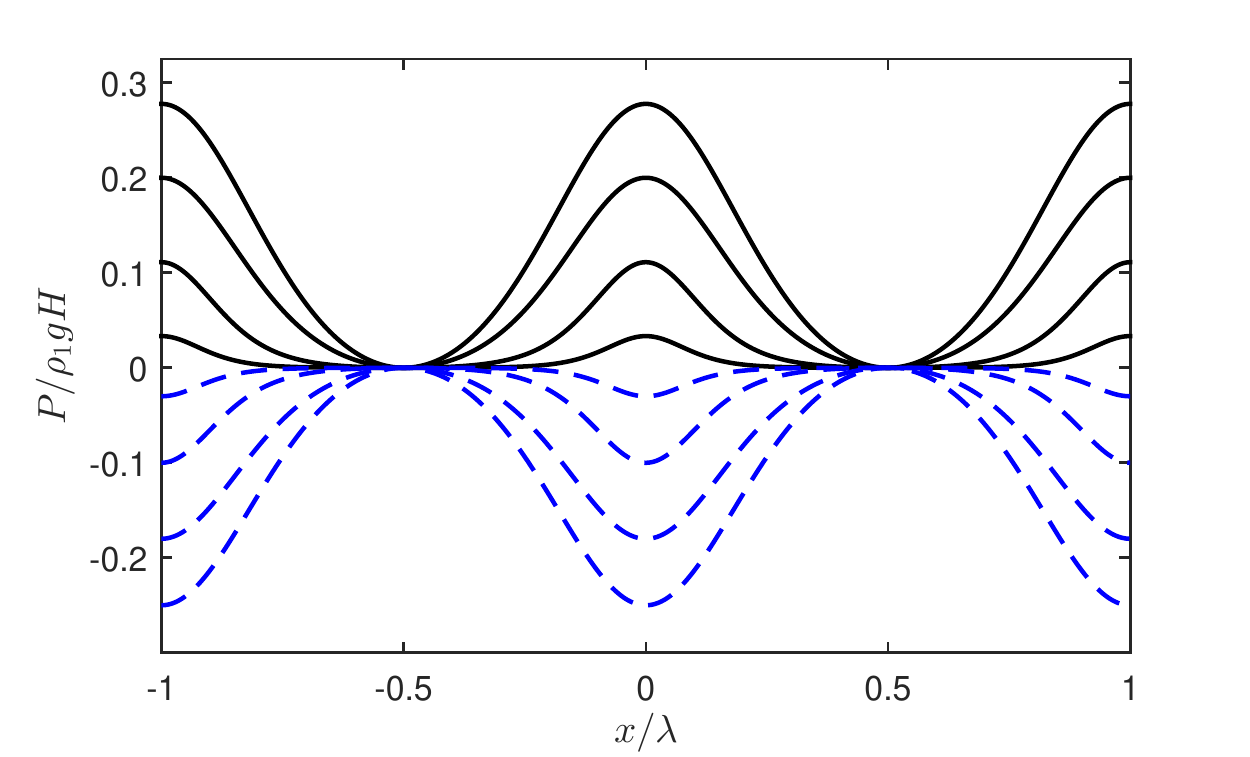} }\newline

\subfigure[~]{\includegraphics[width = 7cm,clip]{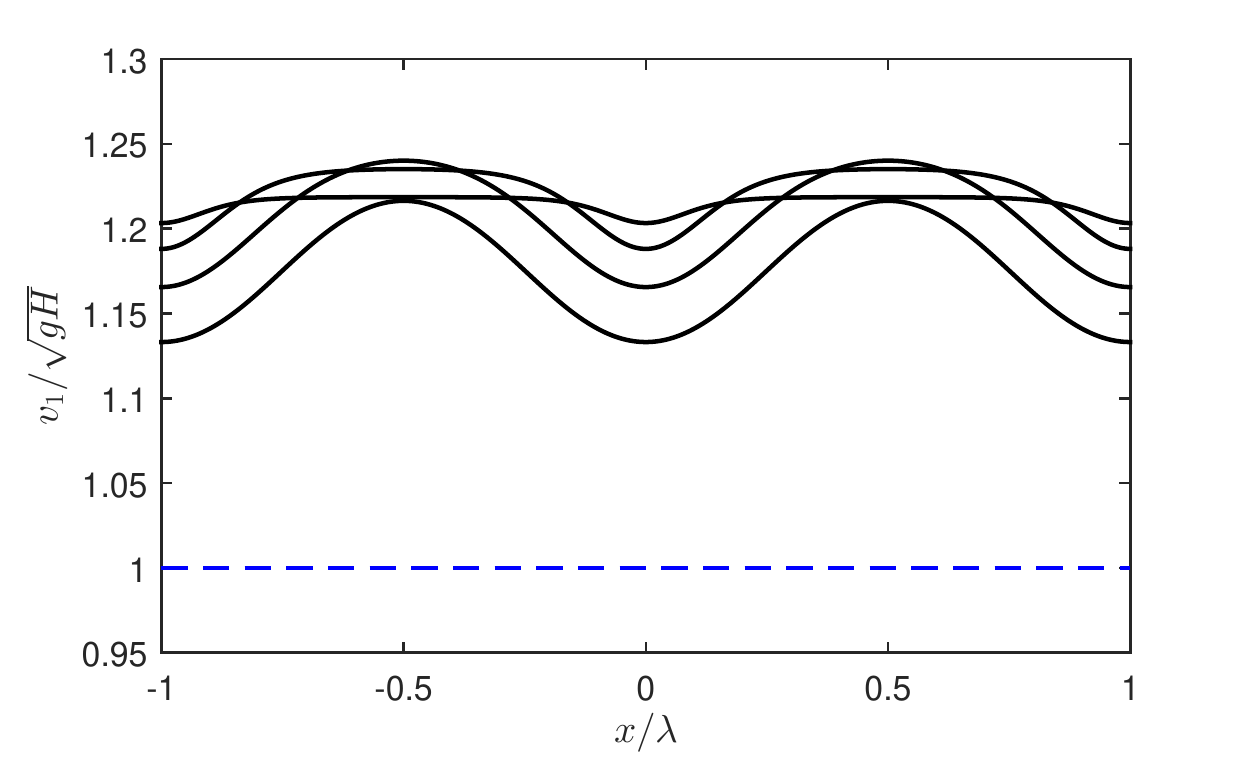} }
\subfigure[~]{\includegraphics[width = 7cm,clip]{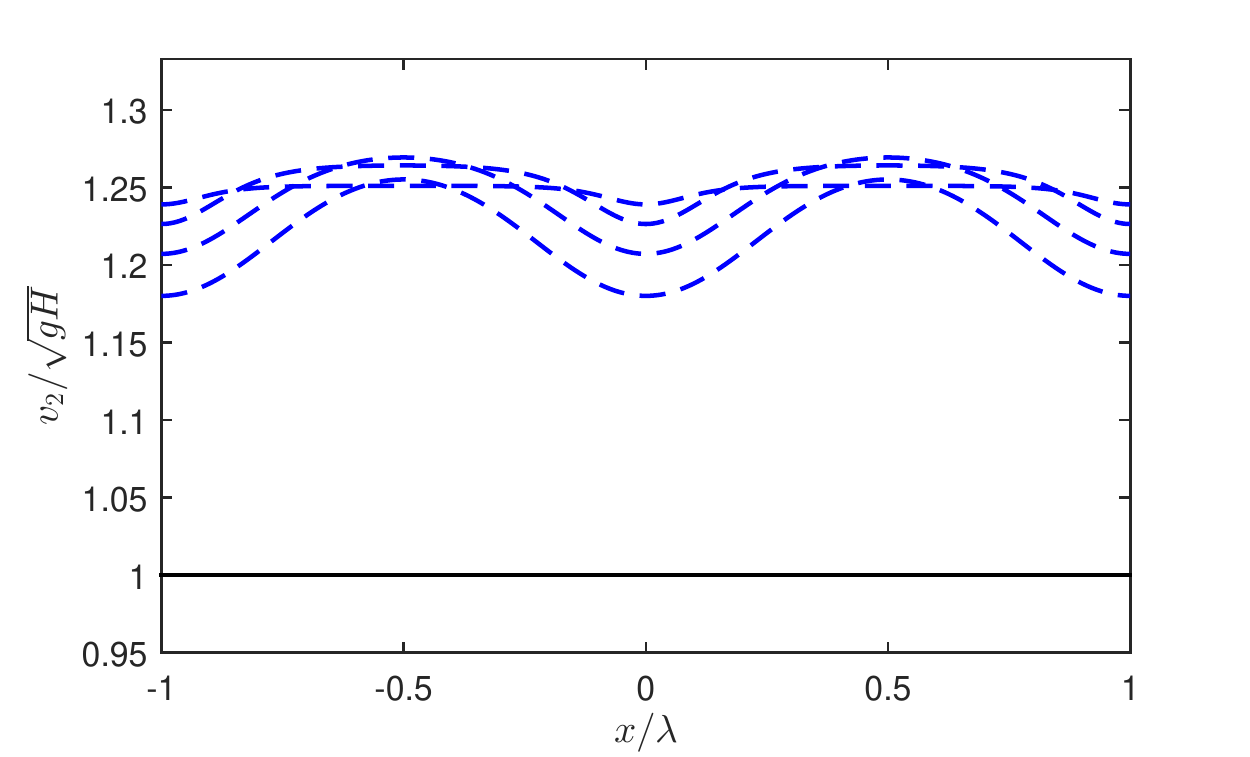} }\newline

\caption{Dimensionless flow parameter curves for the periodic cnoidal wave exact solution family \eqref{eq:trw:snSQ:sol:P}, \eqref{eq:trw:v1v2:solA}, \eqref{eq:trw:v1v2:solB}, \eqref{eq:trw:snSQ:sol:Zeta}. Case 1 curves are shown in solid black, with amplitudes $B_1<0$ (Table \ref{tab:sn2:kB:cases}, rows 1, 3, 5, 7); Case 2 curves are dashed blue, $B_1>0$ (small to large amplitude, Table \ref{tab:sn2:kB:cases}, rows 2, 4, 6, 8). (a): the dimensionless interface displacement; (b) the dimensionless pressure at the interface; (c), (d): the dimensionless average horizontal velocities of the upper and the lower fluid. The actual (dimensional) spatial wavelengths of the presented solutions are not equal; they are given in Table \ref{tab:sn2:kB:cases}.
}
\label{fig:sol:sn2:case1}
\end{center}
\end{figure}

Sample plots of exact solutions $\zeta$, $v_{1}$, $v_{2}$, $P$ of the Choi-Camassa system \eqref{eq:CC}, as functions of dimensionless spatial coordinate $x/\lambda$, are presented in Figure \ref{fig:sol:sn2:case1} for the parameter values
\begin{subequations}\label{eq:fig:ch2:case12:params}
\beq\label{eq:fig:ch2:case12:params:1}
\hat{c}=1, \quad h_1=0.4~$m$, \quad h_2=0.6~$m$, \quad H=1~$m$, \quad g=9.8~$m/s$^2, \quad x_0=t=0,
\eeq
and the density ratio
\beq\label{eq:fig:ch2:case12:paramsS}
S=0.9.
\eeq
\end{subequations}
for a set of values of $k$, $0<k<1$, and $B_1$. The above choice of parameters $\hat{c}, H$ corresponds to the dimensional wave speed $c=\hat{c}\sqrt{gH} \simeq 3.13~\text{m/s}$. Formulas \eqref{eq:trw:snSQ:sol:P}, \eqref{eq:trw:v1v2:solA}, \eqref{eq:trw:v1v2:solB}, \eqref{eq:trw:snSQ:sol:Zeta} are used; in \eqref{eq:trw:v1v2:solA} and \eqref{eq:trw:v1v2:solB}, positive signs are chosen. The four black curves in Figure \ref{fig:sol:sn2:case1} correspond to Case 1 solutions plotted for $B_1<0$, and represent periodic surface depression waves, with constant layer-average horizontal speed of the lower fluid. The dashed blue curves for Case 2 solutions are shown for $B_1>0$, corresponding to surface elevations, and constant layer-average horizontal speed values of the upper fluid. The plotted curves correspond to the values of $(k,B_1)$ pairs and wavelengths given in Table \ref{tab:sn2:kB:cases}. We note that despite of the same values of the amplitude of the elliptic cosine, $|B_1|$, the solutions curves for Case 1 and Case 2 are not symmetric. In particular, the last, dimensional plot in Figure \ref{fig:sol:sn2:case1} shows that the wavelengths of the oscillations are different; this is due to the difference of the expressions for $\alpha_1$ in Case 1 and Case 2, with $k, B_1$ taken the same for both cases.

\begin{figure}[ht]
\begin{center}
\subfigure[~]{\includegraphics[width = 14cm,clip]{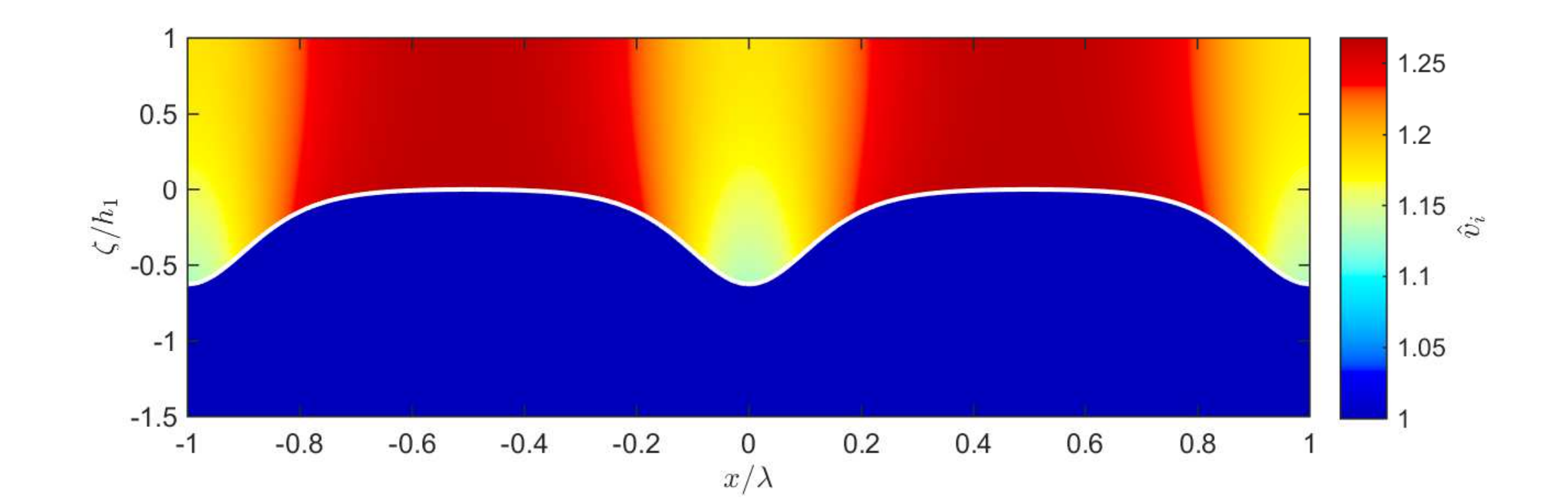} } \newline
\subfigure[~]{\includegraphics[width = 14cm,clip]{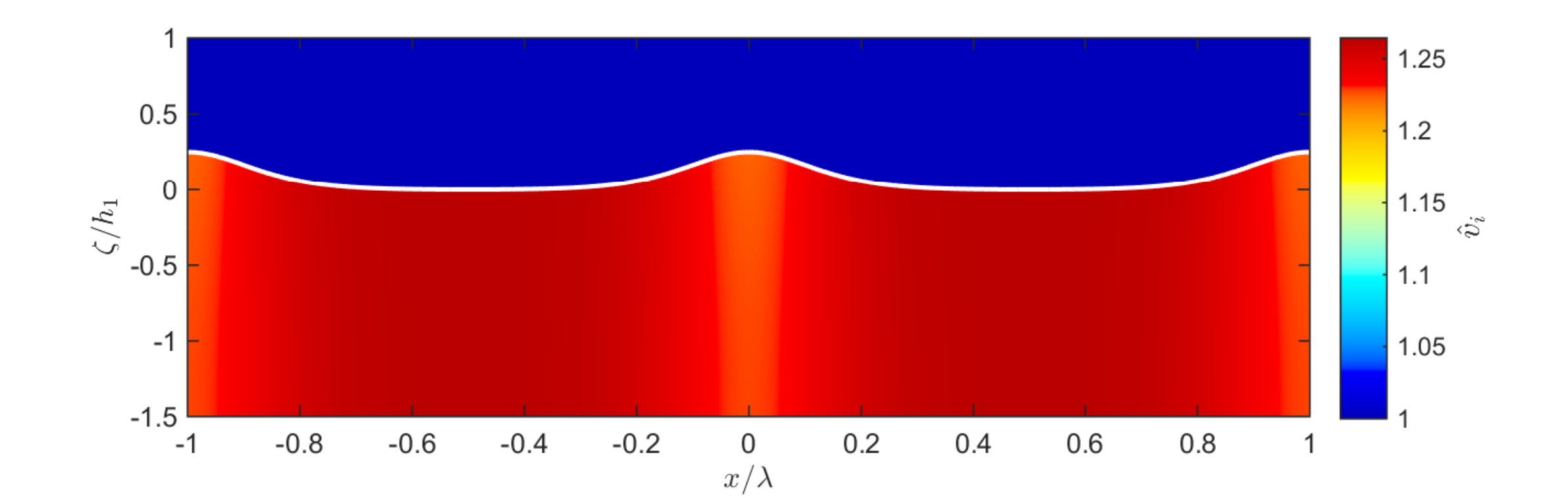} }\newline
\caption{Flood diagrams for the right-propagating cnoidal wave solutions \eqref{eq:th:sol:elliptic:Z:snSQ}, showing dimensionless values of the fluid interface displacement (white curve) and the $(x,z)$-dependent horizontal velocities $u_i$ computed through the  formulas \eqref{eq:ui:thru:vCC}. Figures are given for the solution parameters \eqref{eq:fig:ch2:case12:params} with (a) $k=0.99$, $B_1=-0.25$ for Case 1, and (b) $k=0.99$, $B_1=0.1$ for the Case 2 solution. The corresponding dimensional spatial wavelengths are given in Table \ref{tab:sn2:kB:cases}.
}
\label{fig:sol:sn2:case12:flood}
\end{center}
\end{figure}

Figure \ref{fig:sol:sn2:case12:flood} shows sample flood diagrams for the horizontal velocities $u_i (t, x, z)$ computed through the asymptotic formulas \eqref{eq:ui:thru:vCC} for the cnoidal wave solutions \eqref{eq:trw:snSQ:sol:Zeta} (Cases 1 and 2).

\begin{remark}
Depending on the choice of free parameters listed in Theorem \ref{th:sol:snSQ}, exact solutions of the CC model given by formulas \eqref{eq:th:sol:elliptic:Z:snSQ}, \eqref{eq:trw:snSQ:sol:P}, \eqref{eq:trw:v1v2:solA}, \eqref{eq:trw:v1v2:solB} may or may not satisfy the asymptotic requirement $\epsilon \ll 1$ \eqref{eq:asymp:hL}. Choosing values of $k$ closer to 1, one can unboundedly increase the wavelength $\lambda=L$ \eqref{eq:th:sol:elliptic:Z:snSQ:Wavelength}. Exact solutions of both small and large amplitude exist within the indicated class. Some examples are provided in Table \ref{tab:sn2:kB:cases}, which contains parameter values used to produce Figures \ref{fig:sol:sn2:case1} and \ref{fig:sol:sn2:case12:flood}.
\end{remark}

\begin{table}[H]
  \centering
  \begin{tabular}{|c|c|c|c|c|}
     \hline
     Case & $k$ & $B_1$ & $\lambda$, m & $\epsilon=H/\lambda$ \\ \hline \hline
     1 & 0.9990 & -0.0300 & 15.7055 & 0.0637 \\
     2 & 0.9990 & 0.0300 &  90.4410 & 0.0111 \\\hline
     1 & 0.9900 & -0.1000 & 6.8466 & 0.1461 \\
     2 & 0.9900 & 0.1000 & 22.0327 & 0.0454 \\\hline
     1 & 0.9000 & -0.1800 & 3.2188 & 0.3107 \\
     2 & 0.9000 & 0.1800 & 7.1438  & 0.1400 \\\hline
     1 & 0.8000 & -0.2500 & 1.9146 & 0.5223 \\
     2 & 0.8000 & 0.2500 & 3.1912 & 0.3134 \\\hline
     1 & 0.9900 & -0.2500 &  5.2898 & 0.1890 \\
     \hline
   \end{tabular}
  \caption{Sample exact solution parameters and wavelengths for the exact periodic cnoidal wave solutions \eqref{eq:th:sol:elliptic:Z:snSQ}.}\label{tab:sn2:kB:cases}
\end{table}

%

\subsection{Solitary waves} \label{sec:trw:exact:kn:solitary}


For both Case 1 and Case 2, solitary waves arise from the solution \eqref{eq:th:sol:elliptic:Z:snSQ} for $k=1$, since
${\sn}^2(y, 1)=\tanh^2 y = 1 - \cosh^{-2}\,y$, and
\beq\label{eq:th:sol:elliptic:Z:snSQ:soli}
\hat{Z}(\hat{r})=({B_{1}}+B_{2}) - {B_{1}}\,\cosh^{-2}(\gamma \,\hat{r}).
\eeq
Solutions corresponding to Case 1 describe propagating fluid interface depression waves, whereas Case 2 corresponds to the elevation waves. Coefficient formulas \eqref{eq:th:snSQ:flas1}, \eqref{eq:th:snSQ:flas2} still hold when $k=1$. The solution family \eqref{eq:trw:snSQ:sol:P}, \eqref{eq:trw:v1v2:solA}, \eqref{eq:trw:v1v2:solB}, \eqref{eq:th:sol:elliptic:Z:snSQ:soli} depends on the arbitrary parameters $B_1$, $B_2$, $\hat{c}$, the physical parameters $S=\rho_1/\rho_2$, $h_1$, $h_2$, $g$, and on the sign choice in the average velocity expressions \eqref{eq:trw:v1v2:solA}, \eqref{eq:trw:v1v2:solB}. It is natural to choose $B_2$ according to the formula \eqref{eq:trw:snSQ:sol:Zeta:cn2:B2}:
\beq\label{eq:th:sol:elliptic:Zeta:soli:B1only}
\hat{\zeta}(x,t)=H{B_{1}}\,\cosh^{-2}(\gamma \,\hat{r}(x,t));
\eeq
then $\zeta\to 0$ as $x\to\pm\infty$.

\begin{figure}[ht]
\begin{center}
\subfigure[~]{\includegraphics[width = 7cm,clip]{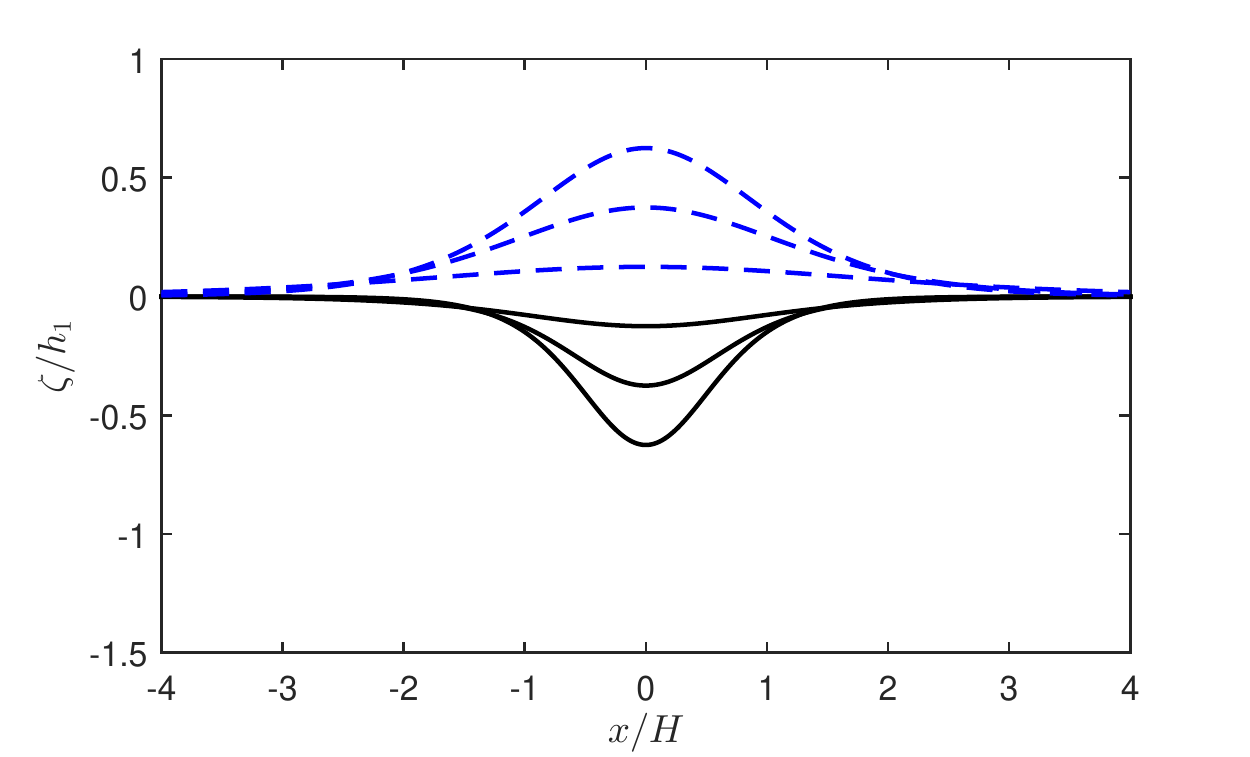} } \subfigure[~]{\includegraphics[width = 7cm,clip]{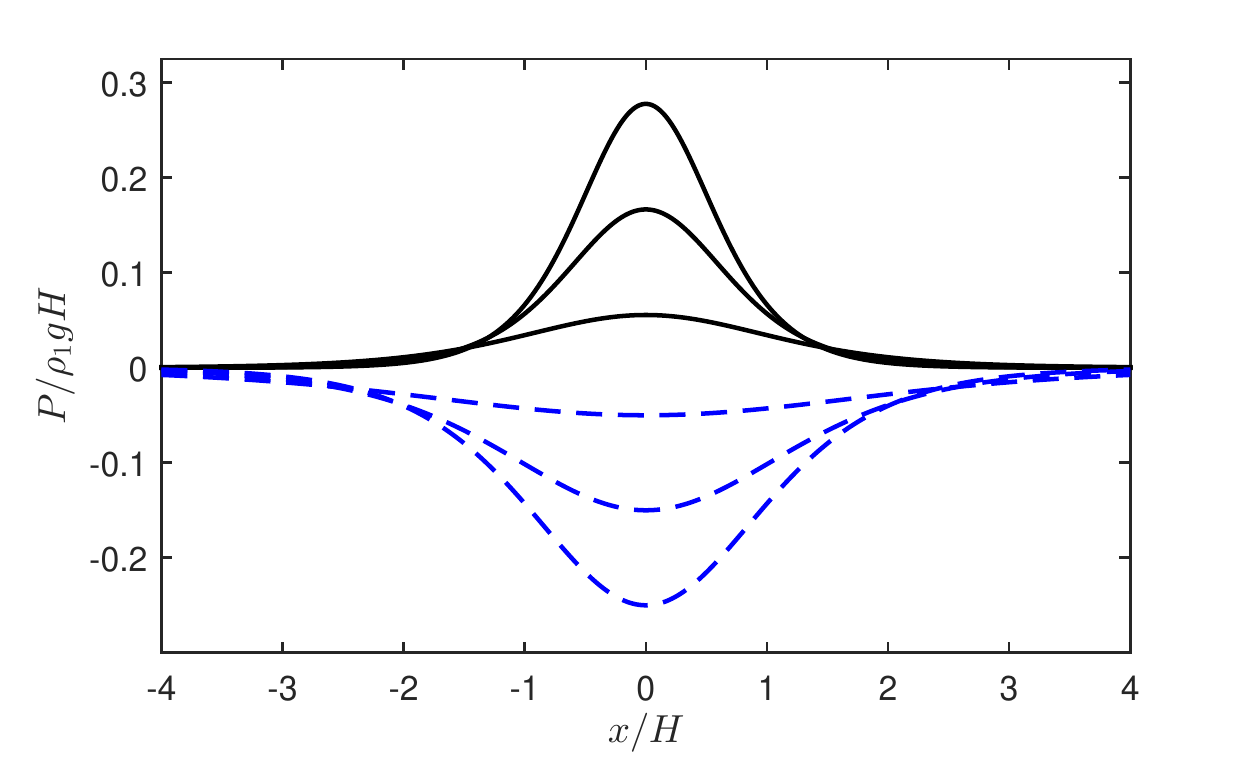} }\newline

\subfigure[~]{\includegraphics[width = 7cm,clip]{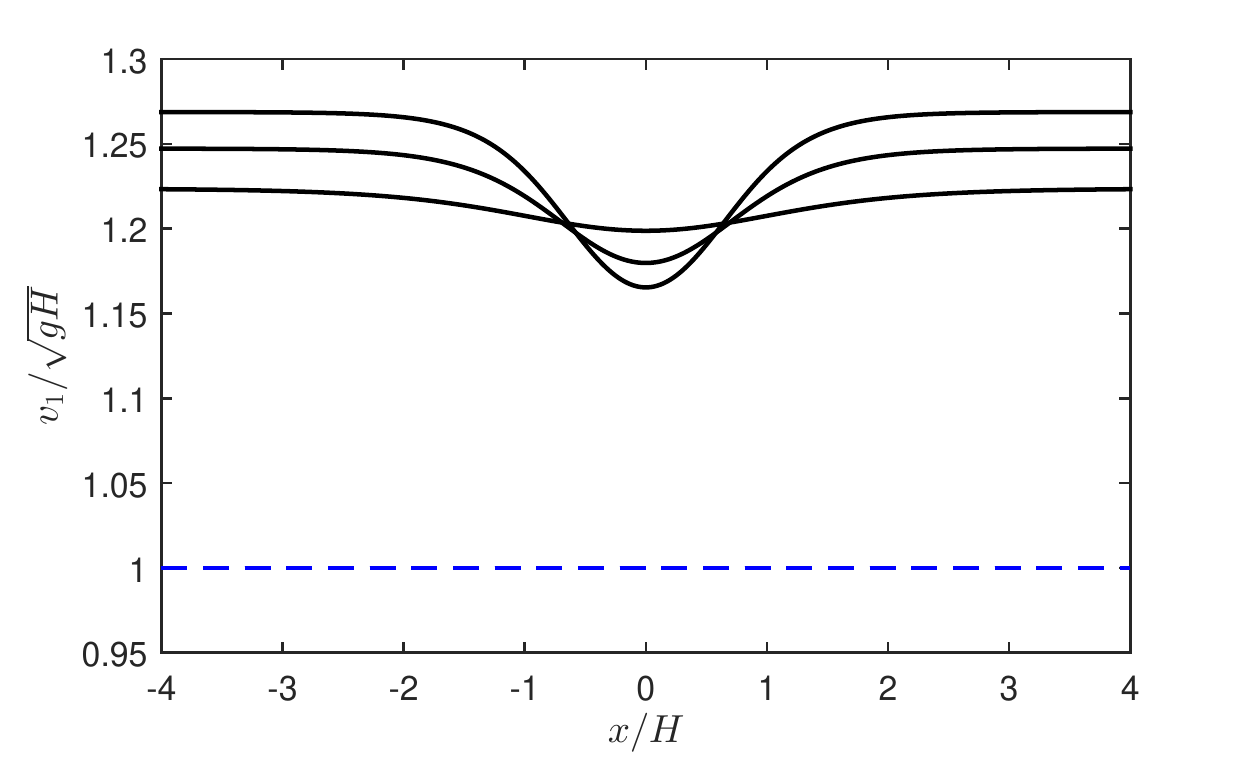} }
\subfigure[~]{\includegraphics[width = 7cm,clip]{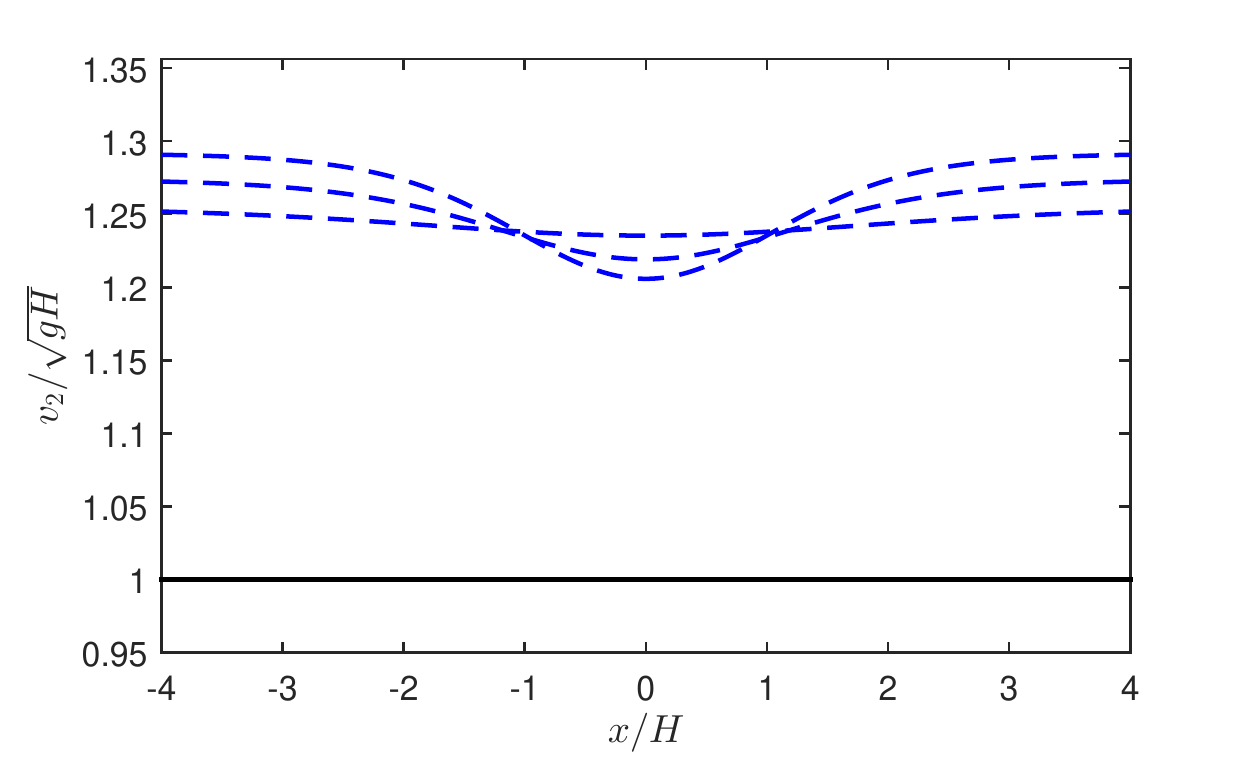} }\newline
\caption{Dimensionless flow parameter curves for the right-propagating solitary wave exact solution families \eqref{eq:trw:snSQ:sol:P}, \eqref{eq:trw:v1v2:solA}, \eqref{eq:trw:v1v2:solB}, \eqref{eq:th:sol:elliptic:Z:snSQ:soli}. Case 1 curves are shown in solid black, with amplitudes $B_1=-0.05, -0.15, -0.25$; Case 2 curves are dashed blue, for $B_1=0.05, 0.15, 0.25$ (small to large amplitude). In this figure, the dimensionless spatial coordinate is the one normalized by the total channel depth: $x/H$.
(a): dimensionless interface displacement; (b) dimensionless pressure at the interface; (c), (d): dimensionless average horizontal velocities of the upper and the lower fluid.
}
\label{fig:sol:sn2:solitons12}
\end{center}
\end{figure}

In Figure \ref{fig:sol:sn2:solitons12}, sample right-propagating depression and elevation-type solitary wave exact  solution profiles are shown for channel/fluid parameters \eqref{eq:fig:ch2:case12:params}, for $B_1=-0.05, -0.15, -0.25$ (Case 1) and $B_1=0.05, 0.15, 0.25$ (Case 2), and $B_2$ given by \eqref{eq:trw:snSQ:sol:Zeta:cn2:B2}. In formulas \eqref{eq:trw:v1v2:solA} and \eqref{eq:trw:v1v2:solB}, positive signs are chosen.

\begin{figure}[ht]
\begin{center}
\subfigure[~]{\includegraphics[width = 14cm,clip]{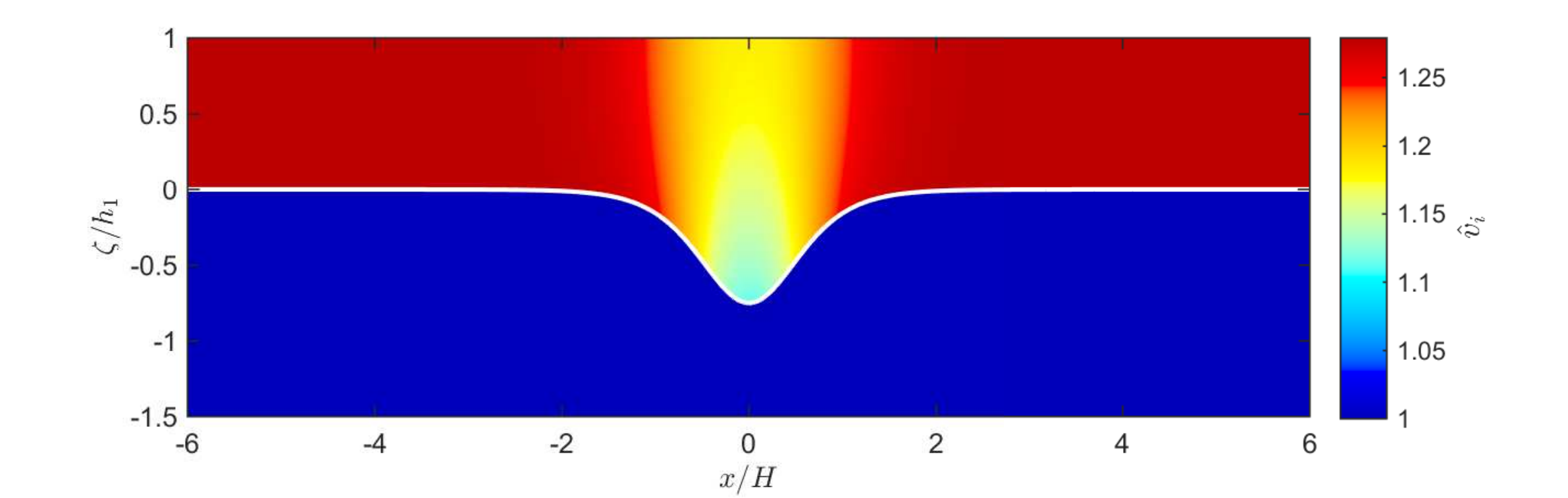} } \newline
\subfigure[~]{\includegraphics[width = 14cm,clip]{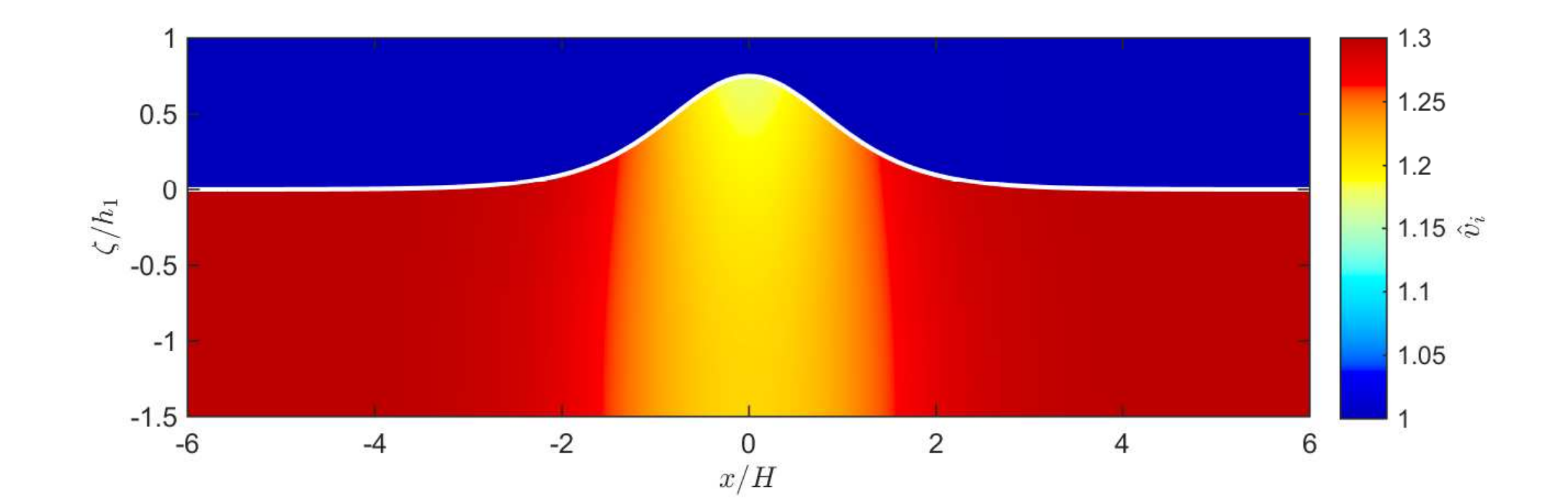} }\newline
\caption{Flood diagrams for the solitary wave solutions \eqref{eq:th:sol:elliptic:Z:snSQ:soli}, showing dimensionless values of the fluid interface displacement (white curve) and the $(x,z)$-dependent horizontal velocities $u_i$ computed through the formulas \eqref{eq:ui:thru:vCC}. Figures are given for the solution parameters \eqref{eq:fig:ch2:case12:params} with (a) $B_1=-0.3$, Case 1; and (b) $B_1=0.3$, Case 2. The spatial coordinate is the one normalized by the total fluid height: $x/H$.
}
\label{fig:sol:sn2:case12:flood:soli}
\end{center}
\end{figure}

As in the periodic case, for the solitary wave solution family, one of the layer-average velocity values is constant. Sample plots of the approximate actual (non-average) horizontal velocity values $u_i(t, x, z)$, calculated through the asymptotic formulas \eqref{eq:ui:thru:vCC} in a moving frame, are shown in flood diagrams in Figure \ref{fig:sol:sn2:case12:flood:soli}.

We are now interested in a relationship between the wave amplitude and the typical wavelength of the exact solitary wave solutions \eqref{eq:th:sol:elliptic:Z:snSQ:soli}. The argument of the hyperbolic cosine is $\gamma \,\hat{r}\sim \gamma x/H$, hence one defines the dimensional wavelength as
\beq\label{eq:soliton:wavelength}
\lambda_{s} = \dfrac{H}{\gamma(B_1,B_2)}.
\eeq
We use $\lambda_{s}^{(1)}$, $\lambda_{s}^{(2)}$ to denote the wavelengths \eqref{eq:soliton:wavelength} arising for Cases 1 and 2 (formulas \eqref{eq:th:snSQ:flas1} and \eqref{eq:th:snSQ:flas2}), respectively.

Denote $R_h=h_1/h_2$. In \cite{koop1981investigation}, for $R_h=5.09$ and the density ratio $S=0.63$, for solitary waves of elevation, experimental measurements of the dimensionless wavelength $\lambda_{s}/{h_2}$ versus the dimensionless wave amplitude $a/{h_2}$ were presented. We consider a similar wavelength-amplitude relationship for the exact solitary wave solutions \eqref{eq:th:sol:elliptic:Zeta:soli:B1only}.

For the elevation-type solitary waves (Case 2, $B_1>0$), let $B_2$ be given by the formula \eqref{eq:trw:snSQ:sol:Zeta:cn2:B2} (or, in general, by $B_2=\const-B_1$), to yield amplitude-independent fluid depths. The elevation amplitude is given by $a=HB_1$. Using \eqref{eq:th:snSQ:flas2}, it is straightforward to show that the dimensionless wavelength-amplitude relationship is given by
\beq\label{eq:soliton:Koop:DimlessLam_of_DimlessB1:elev}
\hat{\lambda}_s^{(2)} = f(q)= \dfrac{2}{\sqrt{3}}\sqrt{1+q^{-1}},
\eeq
were $\hat{\lambda}_{s}^{(2)}=\lambda_{s}^{(2)}/{h_2}$, $q={a}/{h_2}$. Interestingly, the relationship \eqref{eq:soliton:Koop:DimlessLam_of_DimlessB1:elev} is independent of both the fluid density ratio $S$ and the fluid depth ratio $R_h$.

For the depression-type exact solitary wave solutions (Case 1, $B_1<0$), with the same choice of $B_2$, the amplitude is defined as $a=H|B_1|$. The expression for $\lambda_{s}^{(1)}/{h_2}$ as a function of ${a}/{h_2}$ does not turn out elegant, in particular, it is dependent on the fluid depths. However, using the upper fluid depth $h_1$ for non-dimensionalization, and denoting $\hat{\lambda}_s^{(1)}=\lambda_{s}/{h_1}$, $q'={a}/{h_1}$, one arrives at the \emph{same formula} $\hat{\lambda}_s^{(1)}=f(q')$ as \eqref{eq:soliton:Koop:DimlessLam_of_DimlessB1:elev}. This is another manifestation of the partial ``fluid interchange" symmetry mentioned earlier.

\begin{figure}[ht]
\begin{center}
\includegraphics[width = 7cm,clip]{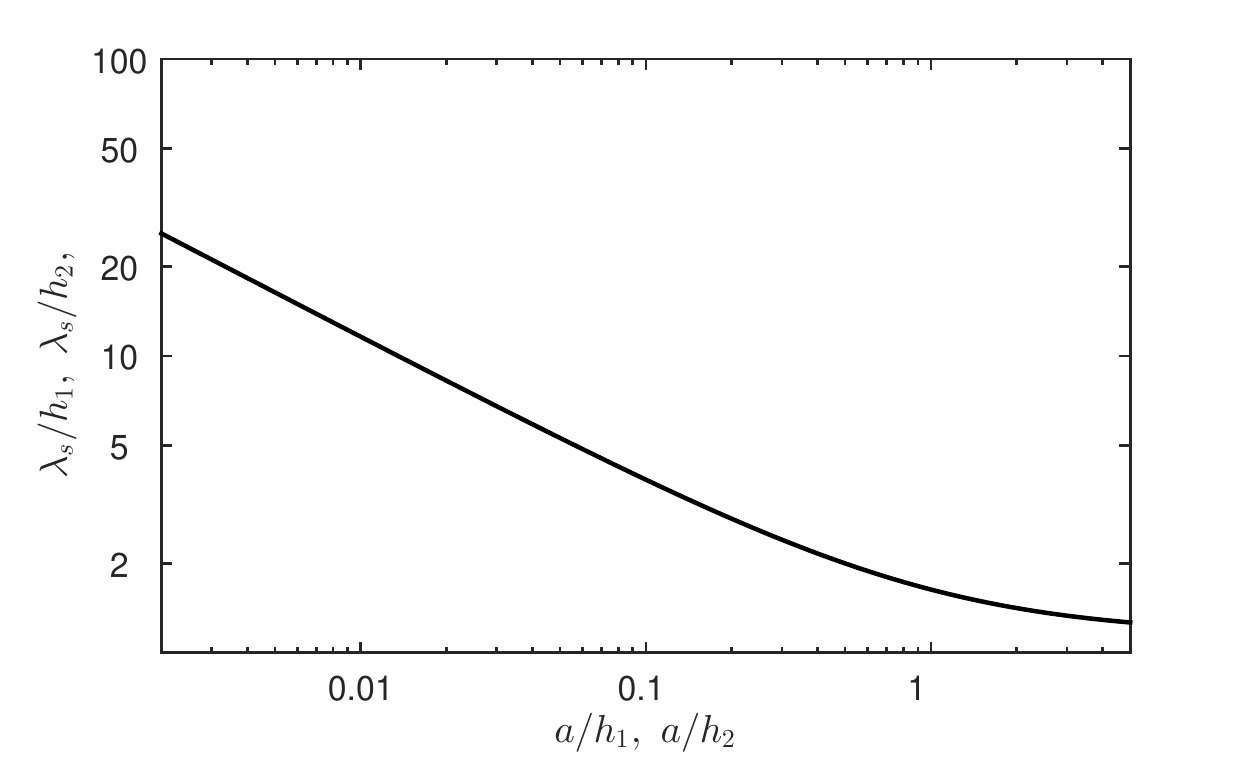}
\caption{The dimensionless wavelength-amplitude relationship $\hat{\lambda}_s= f(q)$ \eqref{eq:soliton:Koop:DimlessLam_of_DimlessB1:elev} for the exact solitary wave solutions \eqref{eq:th:sol:elliptic:Z:snSQ:soli}. Here $q={a}/{h_1}$ and $a=H|B_1|$ for depression waves (Case 1), and $q={a}/{h_2}$, $a=HB_1$ for elevation waves (Case 2).
}
\label{fig:sol:soli:LambdaOfB1}
\end{center}
\end{figure}

In Figure \ref{fig:sol:soli:LambdaOfB1}, a plot of the wavelength-amplitude relationship \eqref{eq:soliton:Koop:DimlessLam_of_DimlessB1:elev} is shown.

\section{Further Periodic and Kink-type Solutions}\label{sec:trw:exact:denom}

Further families of exact solutions of the Camassa-Choi model arise from the traveling wave ODE \eqref{eq:ODE:Z2} as follows.

\begin{theorem} \label{th:sol:sn:denom}
The ODE \eqref{eq:ODE:Z2} admits exact solutions in the form
\beq\label{eq:th:sol:elliptic:sn:denom}
\hat{Z}(\hat{r})=\dfrac{B_{1}}{{\sn}(\gamma \,\hat{r}, k)+B_{2}},
\eeq
for arbitrary constants $B_1, B_2, S$. The remaining constants $\gamma$, $k$, and $\alpha_{1,2}$ are given by the formulas \eqref{eq:ODE:Z2Ai} and any one of the three relationships \eqref{eq:th:Den:flas1}, \eqref{eq:th:Den:flas2}, \eqref{eq:th:Den:flas3} listed in  Appendix \ref{app:Th:Sn:Den}.
\end{theorem}
The above result is also verified by a direct substitution of \eqref{eq:th:sol:elliptic:sn:denom} into the ODE \eqref{eq:ODE:Z2}. One consequently has three families of exact solutions of the dimensionless Choi-Camassa system \eqref{eq:CC:nondim:sys}, each depending on \emph{three arbitrary constant parameters} $B_1, B_2, \hat{c}$, as well as on the arbitrarily prescribed channel/fluid parameters $S=\rho_1/\rho_2$, $h_1$, $h_2$ and the free fall acceleration $g$.

The solution families arising from \eqref{eq:th:sol:elliptic:sn:denom} are regular and physically meaningful when $|B_2|>1$, $0< \hat{Z}(\hat{r}) < 1$. The dimensional fluid interface position $\zeta(x,t)$ and the flow parameters $v_1(t,x)$, $v_2(t,x)$, $P(t,x)$ are found from  \eqref{eq:CC:nondim:Z}, \eqref{eq:CC:nondim:transf}, \eqref{eq:trw:v1v2}, \eqref{eq:trw:P}. They are essentially different from those described in Section \ref{sec:trw:exact:kn:sol}. Ranges of parameters exist that satisfy the asymptotic requirement \eqref{eq:asymp:hL}.

\begin{description} 
  \item[Case 1.] For the coefficient relationship \eqref{eq:th:Den:flas1}, $\alpha_0+\alpha_1=C_2=0$, and hence, similarly to \eqref{eq:trw:v1v2:solA}, the mean velocity of the bottom layer ${v}_2(t,x)= \const$.

  \item[Case 2.] For the relationship \eqref{eq:th:Den:flas2}, $\alpha_0=C_1=0$, which yields  the constant mean velocity of the top layer, ${v}_1(t,x)= \const$ (cf. \eqref{eq:trw:v1v2:solA}).

  \item[Case 3.] For the solution family determined by \eqref{eq:th:Den:flas3}, both mean horizontal velocities are non-constant.
\end{description}

\begin{figure}[ht]
\begin{center}
\includegraphics[width = 7cm,clip]{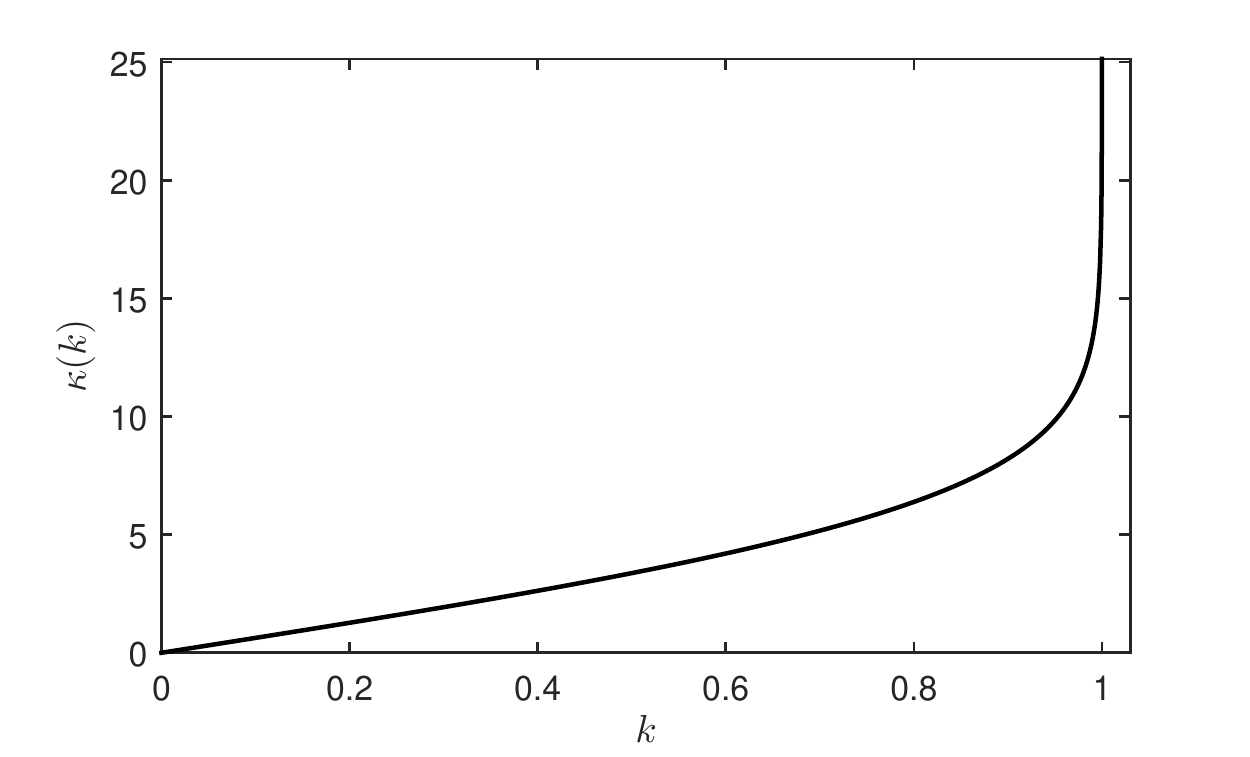}
\caption{The dimensionless wavelength $\hat{\lambda}(k)$ \eqref{eq:th:sol:sn:Den:Wavelength:C3} for the periodic traveling wave solutions \eqref{eq:th:sol:elliptic:sn:denom}, Case 3.
}
\label{fig:sol:sn:Den:Case3:Lamhat:k}
\end{center}
\end{figure}

From the period formula \eqref{eq:sn:T}, the dimensionless and the dimensional $x$-wavelength of exact solutions arising from \eqref{eq:th:sol:elliptic:sn:denom} are computed as follows:
\beq\label{eq:th:sol:sn:Den:Wavelength}
\hat{\lambda}=\dfrac{2\pi}{\gamma\,M(1,\sqrt{1-k^2})},\qquad \lambda=H \hat{\lambda}.
\eeq
For the periodic solutions \eqref{eq:th:sol:elliptic:sn:denom}, the parameters $\gamma$ and $k$ are functions of $B_1$, $B_2$, and do not depend on the density ratio $S$. In particular, for Case 3, one has $k=1/\gamma$, and  \beq\label{eq:th:sol:sn:Den:Wavelength:C3}
\hat{\lambda}(k)=\dfrac{2\pi k}{M(1,\sqrt{1-k^2})},
\eeq
which is plotted in Figure \ref{fig:sol:sn:Den:Case3:Lamhat:k}. The limit $k\to 1^-$, $\hat{\lambda}\to +\infty$ corresponds to the cnoidal-kink wave transition. Periodic and kink-type traveling wave exact solutions are discussed in Sections \ref{sec:trw:exact:denom:per} and \ref{sec:trw:tanh} below.

\subsection{Periodic Solutions with Nonconstant Velocities}\label{sec:trw:exact:denom:per}

As a first illustration, we compute periodic solutions to the CC model \eqref{eq:CC} in the form \eqref{eq:th:sol:elliptic:sn:denom}, \eqref{eq:th:Den:flas3}, that is, in Case 3. Choose the physical constants
\beq\label{eq:fig:cn:denom:case3:params:1}
\hat{c}=1, \quad h_1=3/7~$m$, \quad h_2=4/7~$m$, \quad H=1~$m$, \quad g=9.8~$m/s$^2, \quad x_0=t=0,\quad S=0.9.
\eeq

\begin{table}[H]
  \centering
  \begin{tabular}{|c|c|c|c|c|}
     \hline
     $B_1$ & $B_2$ & $k$ & $\lambda$, m & $\epsilon=H/\lambda$ \\ \hline \hline
     2.3995 & 5 & 0.9950 & 20.4057 & 0.0980 \\
     2.3881 & 5 & 0.8996 &  11.3073 & 0.1769 \\
     2.3037 & 5 & 0.6000 &  5.5882 & 0.3579 \\
     \hline
   \end{tabular}
    \caption{Sample exact solution parameters and wavelengths for the exact periodic cnoidal wave solutions \eqref{eq:th:sol:elliptic:sn:denom}.}\label{tab:sol:snDen:case3}
\end{table}

\begin{figure}[ht]
\begin{center}
\subfigure[~]{\includegraphics[width = 7cm,clip]{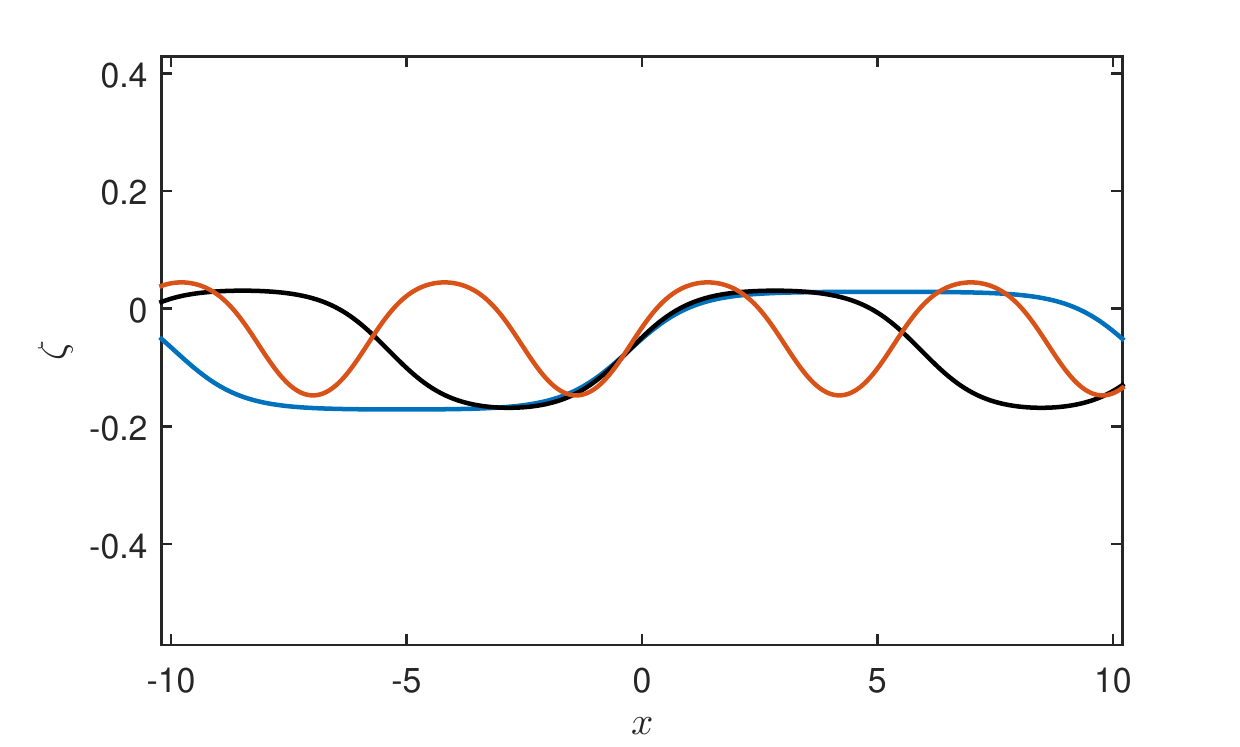} } \subfigure[~]{\includegraphics[width = 7cm,clip]{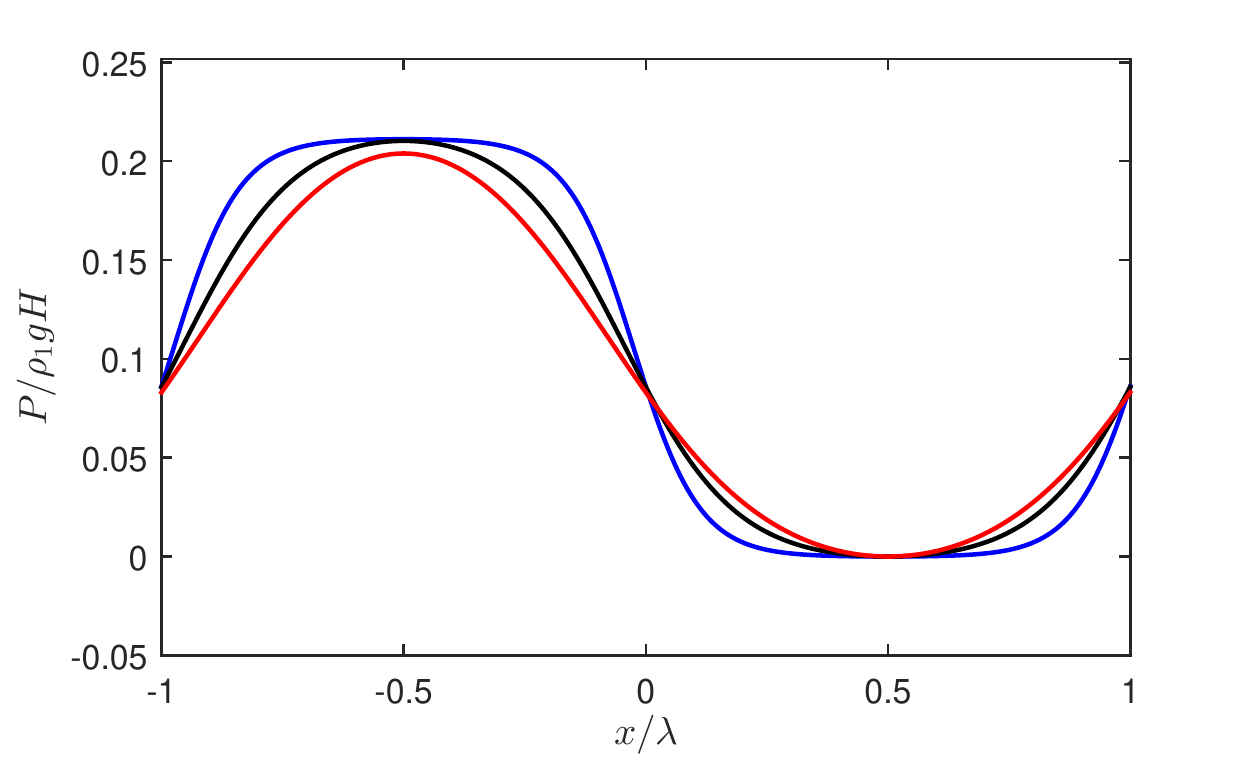} }\newline

\subfigure[~]{\includegraphics[width = 7cm,clip]{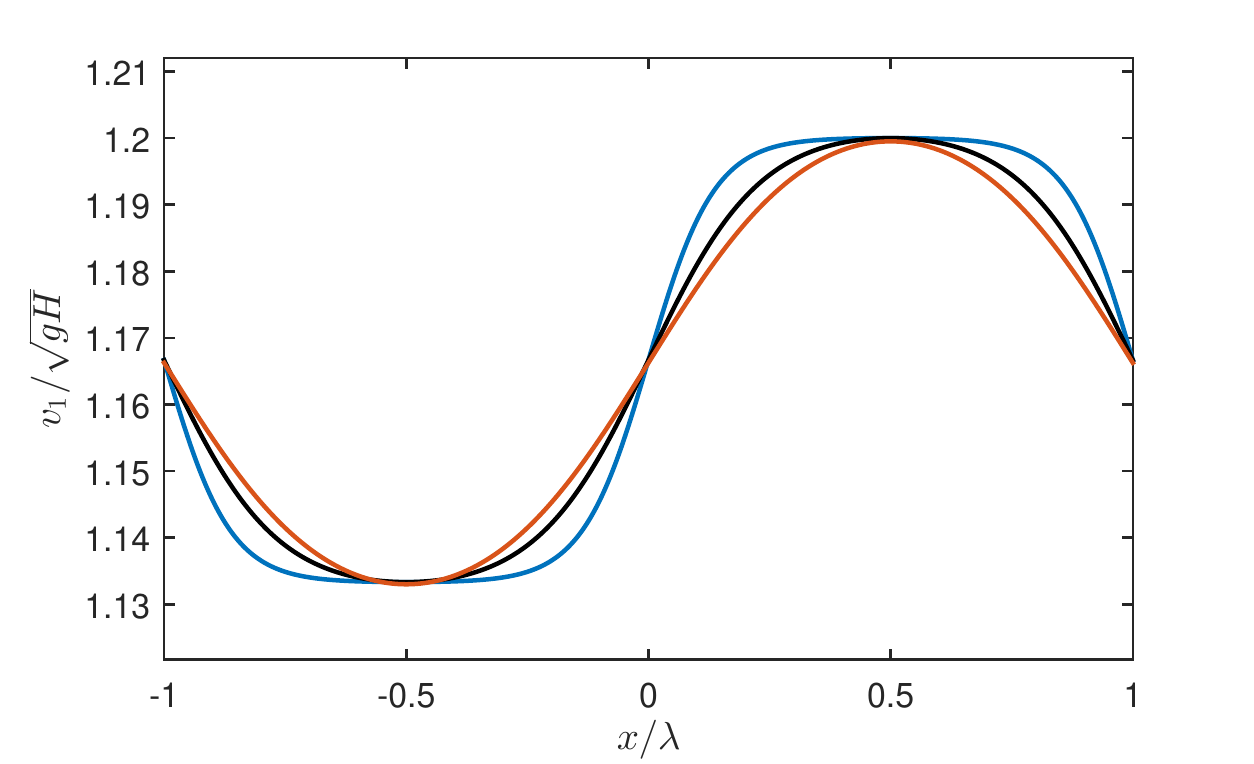} }
\subfigure[~]{\includegraphics[width = 7cm,clip]{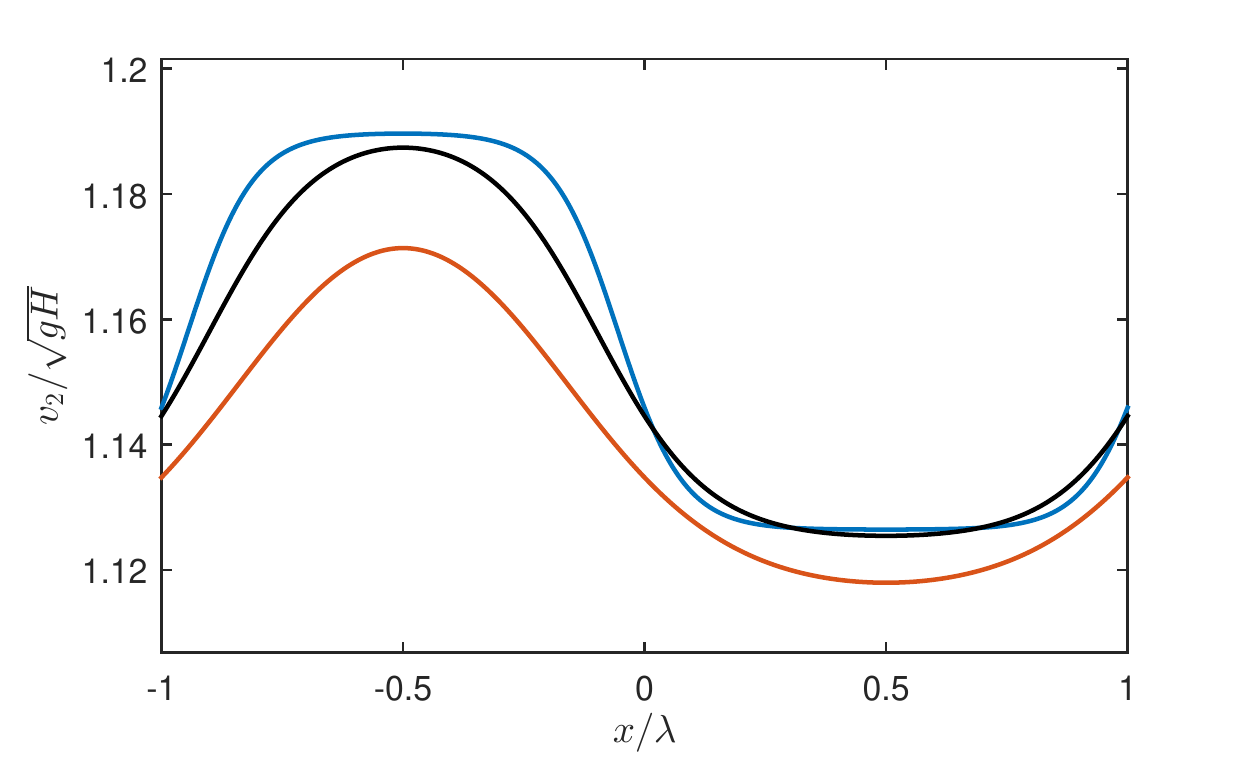} }\newline

\caption{Sample flow parameter curves for the exact periodic solutions \eqref{eq:th:sol:elliptic:sn:denom} of the Camassa-Choi model, Case 3, in the case of right-propagating waves. Curve colors black, blue, and red correspond to the tree rows of Table \ref{tab:sol:snDen:case3}.  (a): dimensional interface displacement; (b) dimensionless pressure at the interface; (c), (d): dimensionless average horizontal velocities of the upper and the lower fluid.
}
\label{fig:sol:snDen:case3}
\end{center}
\end{figure}

\begin{figure}[ht]
\begin{center}
\includegraphics[width = 14cm,clip]{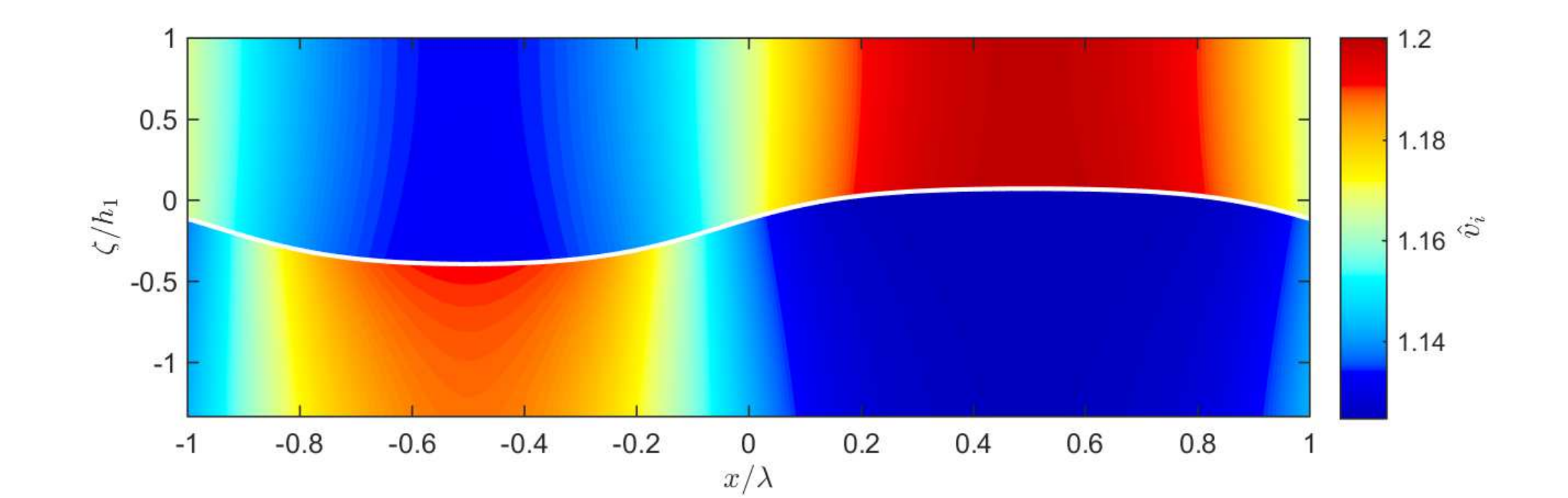} 
\caption{A flood diagram for one period of a right-propagating cnoidal wave solution \eqref{eq:th:sol:elliptic:sn:denom}, showing dimensionless values of the fluid interface displacement (white curve) and the $(x,z)$-dependent horizontal velocities $u_i$ computed through the formulas \eqref{eq:ui:thru:vCC}, for the solution parameters in the second row of Table \ref{tab:sol:snDen:case3}.
}
\label{fig:sol:snden:flood}
\end{center}
\end{figure}

Solution curves in Figure \ref{fig:sol:snDen:case3} are plotted for a set of arbitrary constants $B_1$, $B_2$ and the resulting values of $k$ and the spatial wavelength $\lambda$ are given in Table \ref{tab:sol:snDen:case3}. For the chosen sample parameters, surface wave amplitudes are rather similar, therefore, a dimensional plot of the fluid interface displacement is shown.  The flood diagram in Figure \ref{fig:sol:snden:flood} shows a snapshot the horizontal velocity values $u_i(t, x, z)$ \eqref{eq:ui:thru:vCC} for the parameters in the second row of Table \ref{tab:sol:snDen:case3}. Figures \ref{fig:sol:snDen:case3}, \ref{fig:sol:snden:flood} were produced under the positive sign choice for both average velocities in \eqref{eq:trw:v1v2:alpha}.

\subsection{Exact Kink/Anti-Kink Solutions of the Choi-Camassa Equations}\label{sec:trw:tanh}

Since ${\sn}(y, 1)=\tanh y$, one readily constructs exact kink- and anti-kink-type solutions of the Choi-Camassa PDE system \eqref{eq:CC}
\beq\label{eq:th:sol:tanh:sol}
\hat{Z}(\hat{r})=\dfrac{B_{1}}{{\tanh}(\gamma \,\hat{r})+B_{2}}
\eeq
by setting $k=1$ in formulas \eqref{eq:th:sol:elliptic:sn:denom}, \eqref{eq:th:Den:flas1}, \eqref{eq:th:Den:flas2}, \eqref{eq:th:Den:flas3}. Physically meaningful solutions exist, satisfying, in particular, the condition $0< \hat{Z}(\hat{r}) < 1$.

As an illustration, we consider Case 3, and the coefficient formulas \eqref{eq:th:Den:flas3} with $k=1$.  This yields, in particular, the following relationships between an arbitrary constant $B_1$ and other solution parameters:
\beq\label{eq:th:sol:tanh:const:flasGen} 
\barr
B_2^2-2 B_1 B_2-1=0, 
\qquad \gamma^2=\dfrac{3}{B_1^2},\qquad \alpha_1=0,\\[3ex]
\alpha_0=\dfrac{A_4 B_1^2(2B_2-B_1)}{12(6 B_1^2B_2+3B_1+2B_2)}.
\earr
\eeq
The relationships \eqref{eq:th:sol:tanh:const:flasGen} lead to physical solutions (other relationships exist, in particular, other admissible forms of $B_2$, leading to singular solutions).
From \eqref{eq:th:sol:tanh:const:flasGen}, $B_2=B_1\pm\sqrt{B_1^2+1}$; regular solutions arise with the positive sign choice when $B_1>0$ and the negative sign choice when $B_1<0$. The dimensional amplitude and the characteristic wavelength of the interface displacement for the kink/anti-kink solutions are readily computed from \eqref{eq:CC:nondim:Z}, \eqref{eq:th:sol:tanh:sol}, \eqref{eq:th:sol:tanh:const:flasGen}, and are given by
\beq\label{eq:th:sol:sn:Den:Wavelength:tanh}
a=H|B_2|^{-1}, \qquad \lambda=\dfrac{H}{\gamma}=\dfrac{H|B_1|}{\sqrt{3}}.
\eeq


We note that for \eqref{eq:th:sol:tanh:const:flasGen}, in the limiting case $B_1\to\pm \infty$, the larger root $B_2\simeq 2B_1$, and the kink/anti-kink solutions tend to a constant: $\hat{Z}(\hat{r})\to {1}/{2}$,
corresponding to an equilibrium situation in a channel with equal fluid layer thicknesses $h_1=h_2$.

\begin{figure}[ht]
\begin{center}
\subfigure[~]{\includegraphics[width = 7cm,clip]{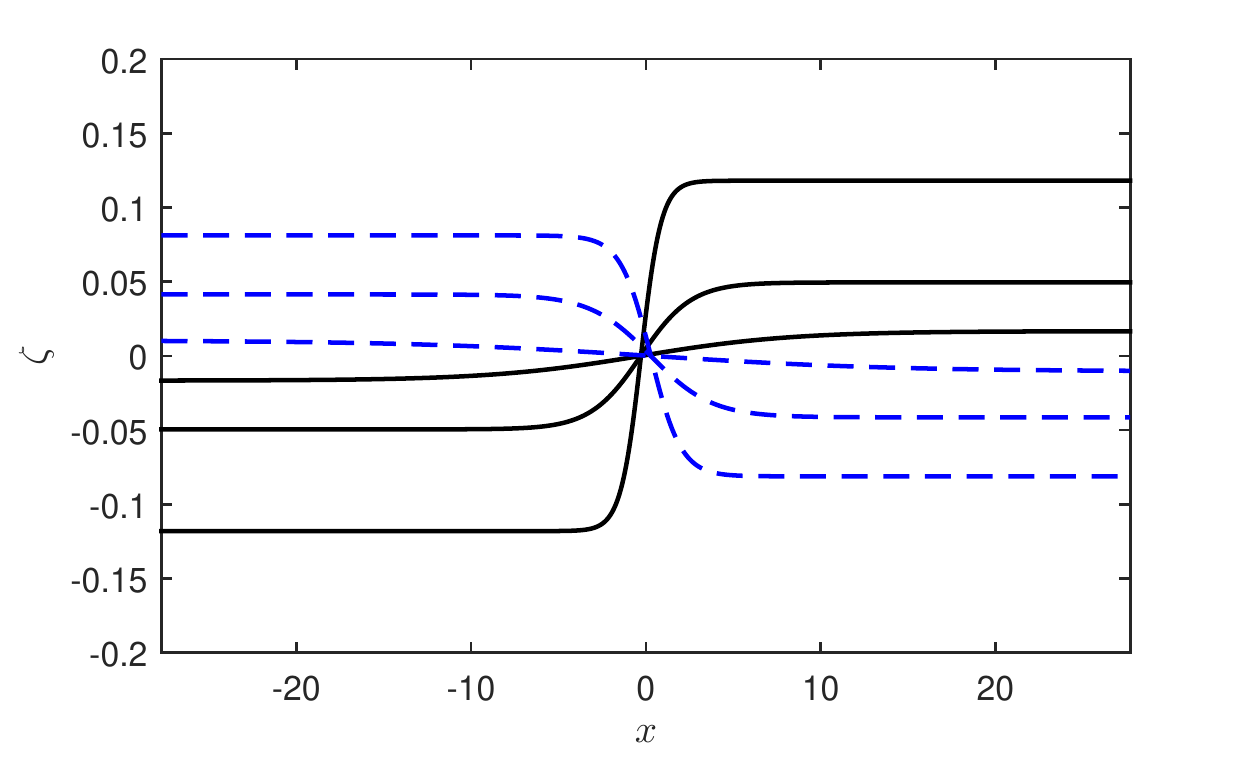} } \subfigure[~]{\includegraphics[width = 7cm,clip]{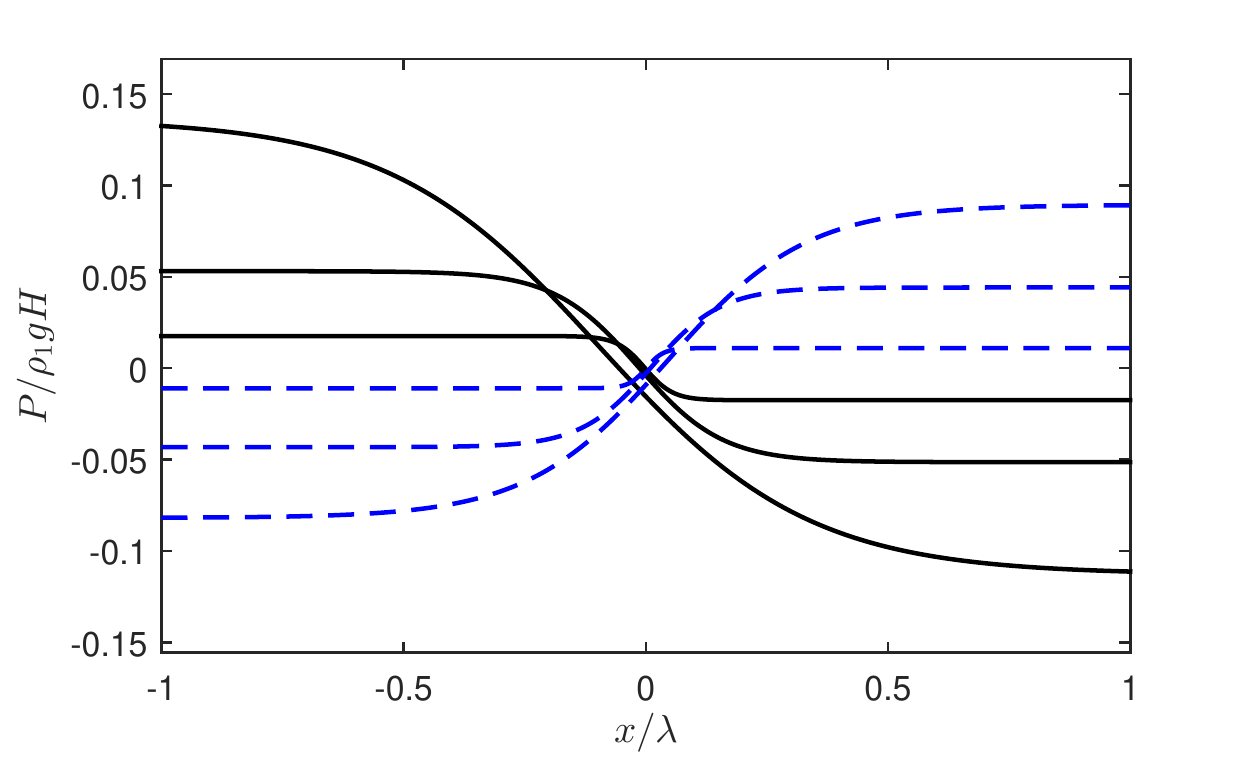} }\newline

\subfigure[~]{\includegraphics[width = 7cm,clip]{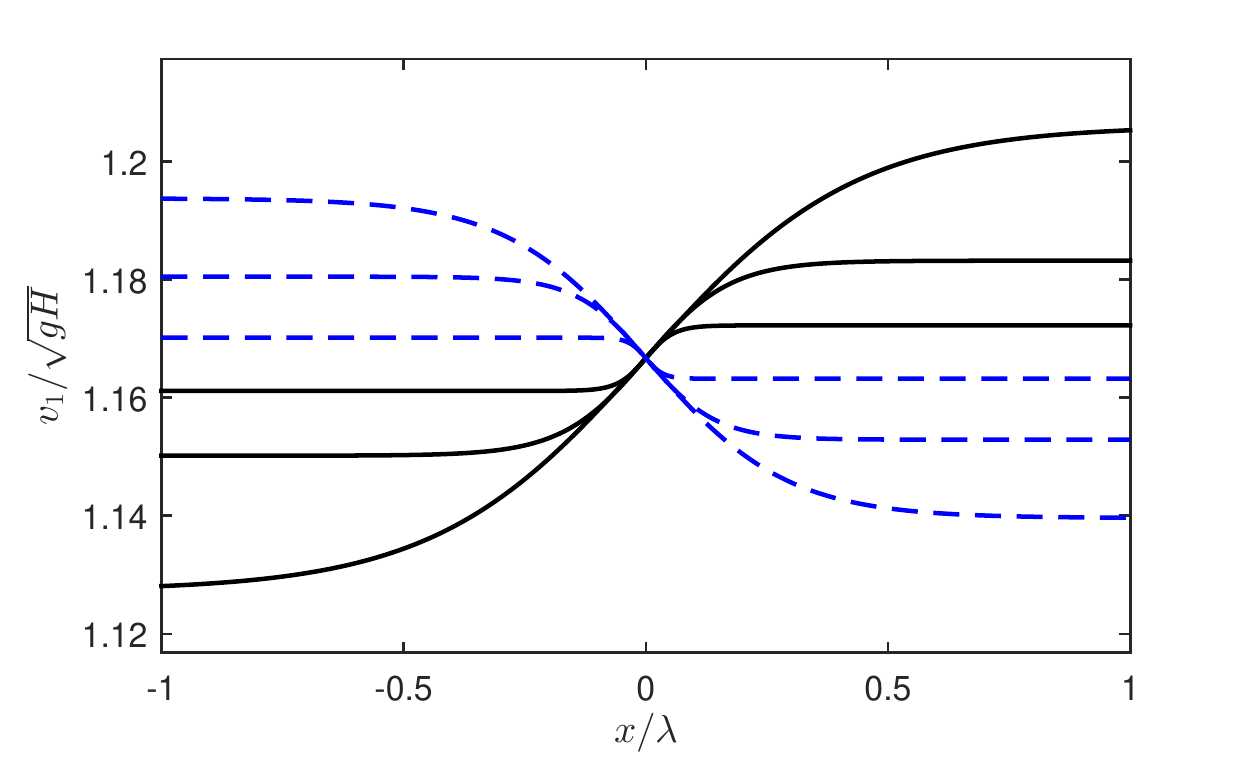} }
\subfigure[~]{\includegraphics[width = 7cm,clip]{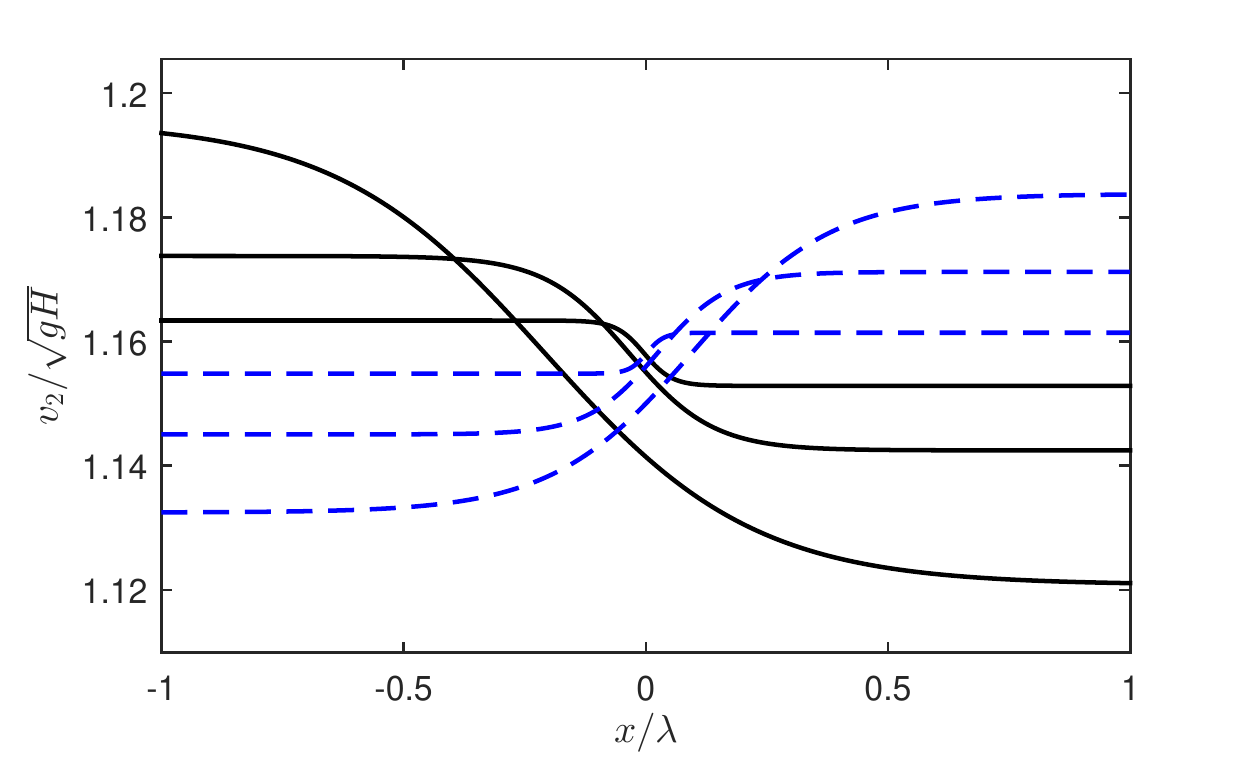} }\newline

\caption{Sample flow parameter curves for right-propagating kink/anti-kink solutions \eqref{eq:th:sol:tanh:sol} of the Camassa-Choi model. Black solid curves (large to small amplitude) correspond to the first tree rows of Table \ref{tab:sol:snDen:Kink} (kink solutions). Blue dashed curves (large to small amplitude) correspond to the rows 4-6 of Table \ref{tab:sol:snDen:Kink} (anti-kink solutions).  (a): dimensional interface displacement; (b) dimensionless pressure at the interface; (c), (d): dimensionless layer-average horizontal velocities of the upper and the lower fluid.
}
\label{fig:sol:TanhDen}
\end{center}
\end{figure}

\begin{figure}[ht]
\begin{center}
\includegraphics[width = 14cm,clip]{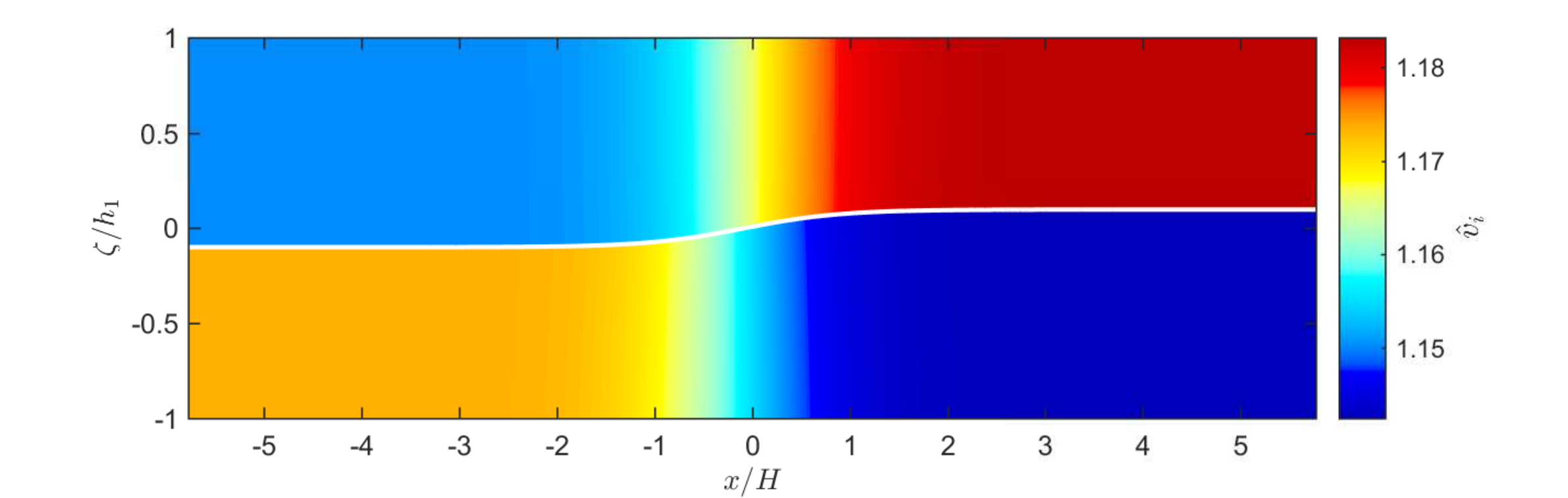} 
\caption{A flood diagram for the kink solution \eqref{eq:th:sol:tanh:sol}, showing dimensionless values of the fluid interface displacement (white curve) and the $(x,z)$-dependent horizontal velocities $u_i$ computed through the formulas \eqref{eq:ui:thru:vCC} for the parameters listed in \eqref{eq:fig:tanh:params} and the second row of Table \ref{tab:sol:snDen:Kink}.
}
\label{fig:sol:TanhDen:flood}
\end{center}
\end{figure}

Plots of sample curves of right-propagating  kink-type exact solutions, for dimensionless parameters given in Table \ref{tab:sol:snDen:Kink} and physical constants
\beq\label{eq:fig:tanh:params}
\hat{c}=1, \quad h_1=h_2=0.5~$m$, \quad H=1~$m$, \quad g=9.8~$m/s$^2, \quad x_0=t=0,\quad S=0.9,
\eeq
and the flood velocity plot corresponding to the second row of Table \ref{tab:sol:snDen:Kink}, are shown in Figures \ref{fig:sol:TanhDen}, \ref{fig:sol:TanhDen:flood}, for the choice of the positive sign of $\gamma$ in \eqref{eq:th:sol:tanh:const:flasGen}.

\begin{table}[H]
  \centering
  \begin{tabular}{|c|c|c|c|c|}
     \hline
     $B_1$ & $B_2$ & $a$ & $\lambda$ & $\epsilon=H/\lambda$\\ \hline \hline
     2 &   4.2361  & 0.8660 & 1.1547 & 0.8660 \\
     5 &  10.0990 & 0.3464 &  2.8868  &0.3464\\
     15 & 30.0333  &  0.1155 & 8.6603 &0.1155\\
     -3 &   -6.1623   & 0.5774 & 1.7321& 0.5774\\
     -6 &  -12.0828 & 0.2887 & 3.4641 &0.2887\\
     -24 &  -48.0208  & 0.2887 & 13.8564 &0.0722\\
     \hline
   \end{tabular}
    \caption{Sample exact solution parameters for the kink/anti-kink exact solutions \eqref{eq:th:sol:tanh:sol}.}\label{tab:sol:snDen:Kink}
\end{table}

\section{Conclusions and Discussion}\label{sec:end}

The Choi-Camassa two-fluid model \eqref{eq:CC} is a nonlinear (1+1)-dimensional asymptotic approximation of the (2+1)-dimensional system of Euler equations and the interfacial conditions between two incompressible stratified fluids of different depth and density in a horizontal channel; the asymptotic assumption is the smallness of the fluid depth/characteristic length ratio. Mathematically, the model is given by a system of four nonlinear partial differential equations for the unknown fluid interface displacement, two layer-average horizontal velocities, and pressure. The PDE system involves mixed space-time third derivatives; it is not a `normal' system of equations in the sense of \cite{OlverBk}.

The CC model \eqref{eq:CC} depends on five physical constitutive parameters \eqref{eq:CC:PhysParam}. In Section \ref{sec:dimless}, a dimensionless form \eqref{eq:CC:nondim:sys} of the model is derived, involving a single dimensionless density ratio parameter \eqref{eq:S:dens:ratio}. Due to the complexity of the nonlinear model, closed-form solutions of generic initial-boundary value problems for the system \eqref{eq:CC} or \eqref{eq:CC:nondim:sys} are not available.

The original CC system and its dimensionless version \eqref{eq:CC:nondim:sys} admit space- and time-translation symmetries, thus allowing for a travelling wave solution ansatz. With the help of systematically calculated integrating factors, the four PDEs in this ansatz reduce to an unusual single first-order nonlinear dimensionless ODE \eqref{eq:ODE:Z2} with a rational polynomial function in the right-hand side. Such a class of ODEs has not been extensively studied in literature; the implicit general solution can be clearly written through an integral, whereas no formula for an explicit general solution is known to date.  The ODE \eqref{eq:ODE:Z2} describes bidirectional travelling wave profiles of the fluid interface displacement; the layer-averaged velocities are consequently computed though the formulas \eqref{eq:trw:v1v2:alpha}, and the pressure is found from \eqref{eq:trw:P}.

In the current work, families of exact physically relevant traveling wave solutions of the CC model were presented, arising from special solutions of the ODE \eqref{eq:ODE:Z2}. These multi-parameter families hold for wide ranges of physical fluid/channel parameters; they include the elevation and depression solitary wave, kink/ani-kink, and periodic traveling waves. Given by closed-form explicit expressions, the exact solutions elucidate some essential features of the model.

\begin{enumerate}
  \item In Section \ref{sec:trw:exact:kn:sol}, a family of cnoidal wave-type solutions \eqref{eq:th:sol:elliptic:Z:snSQ}, depending on eight arbitrary constant parameters, is presented; these parameters are the frequency parameter $k$, the wave amplitude and displacement $B_1, B_2$, the traveling wave speed $\hat{c}$, the channel/fluid constants $S=\rho_1/\rho_2, h_1, h_2$, and the free fall acceleration $g$. The family contains periodic solutions of an arbitrary wavelength, as well as solitary wave-type solutions \eqref{eq:th:sol:elliptic:Z:snSQ:soli} corresponding to the infinite wavelength limit. In particular, both depression and elevation waves arise, for wide ranges of fluid density ratios $S$ and channel depth parameters. All solutions are given by explicit formulae. The wavelength \eqref{eq:th:sol:elliptic:Z:snSQ:Wavelength} of the periodic cnoidal solutions depends on the wave shape parameters $k, B_1, B_2$. For the solitary wave solution, the wavelength is a function of the amplitude, and is determined by the expression \eqref{eq:soliton:Koop:DimlessLam_of_DimlessB1:elev}; the amplitude exponentially decreases at infinity, matching the behaviour of solitary wave solutions of \cite{choi1999fully}. For all exact solutions of Section \ref{sec:trw:exact:kn:sol}, one of the layer-average fluid velocities ($v_1$ for elevation waves, and $v_2$ for depression waves) has a constant value.

  \item A different family of exact periodic solutions of the CC model, also given by explicit expressions involving elliptic integrals, follows from solutions \eqref{eq:th:sol:elliptic:sn:denom} to the ODE \eqref{eq:ODE:Z2} (Section \ref{sec:trw:exact:denom}, Theorem \ref{th:sol:sn:denom}). The exact solutions involve arbitrary constant parameters $B_1, B_2, \hat{c}$, and again hold for an arbitrary choice of the physical constants $S=\rho_1/\rho_2, h_1, h_2, g$. Examples of periodic solutions are resented where, unlike the first family, neither of the layer-average fluid velocities vanishes. In the infinite wavelength limit, this solution family yields exact kink/anti-kink (front-type) solutions involving a hyperbolic tangent.
\end{enumerate}

The exact explicit solutions computed in the current contribution are given by relatively simple expressions involving well-studied elliptic integrals. The correctness of the solutions was verified explicitly by substitutions into the full Choi-Camassa PDE system \eqref{eq:CC}. Both solution families describe left- and right-propagating waves, depending on the choice of the sign of the wave speed $\hat{c}$. Additionally, there is a freedom in both families corresponding to independent choice of the sign in velocity formulas \eqref{eq:trw:v1v2:alpha}. Our solutions generally compare well with semi-numerical ones presented in \cite{choi1999fully,camassa2010fullyPer}, though some aspects of the latter remain unclear. In particular, in the numerical computations of solitary waves in \cite{choi1999fully} it is not clear how the Figure 4, for example, was obtained. The first-order ODE solved numerically through a finite-difference method requires one initial condition, which cannot be the zero slope condition at the origin. If the latter was indeed attained by a shooting-type method, then the graphs in Figure 4 of \cite{choi1999fully} can be interpreted as kink/anti-kink-type solutions joined smoothly together to form a ``solitary wave" profile.

An interesting feature of the traveling wave solutions of Section \ref{sec:trw:exact:kn:sol} is the identically constant value of the layer-average velocity one of the fluids. It is of interest to compare this with experimental data. Actual average velocity values of periodic and solitary wave solutions were discussed neither in the work \cite{choi1999fully,camassa2010fullyPer} nor in the experimental paper \cite{koop1981investigation}.

Another aspect of solitary wave-type solutions discussed in \cite{choi1999fully} is a ``critical depth ratio" $h_1/h_2= (\rho_1/\rho_2)^{1/2}$. In particular, the solitary waves computed in \cite{choi1999fully} were reported to not exist when the depth ratio is critical, and to correspond to waves of elevation and depression for supercritical and subcritical depth ratios, respectively. For the exact solitary wave solutions computed in the current work, the fluid interface elevation \eqref{eq:th:sol:elliptic:Z:snSQ:soli} is independent of the depth ratio, and hence no critical ratio arises. Both depression- and elevation-type solitons exist for wide ranges of amplitude, frequency, and channel/fluid parameters.

Future work directions and open questions include the stability study of the presented solutions, in particular, in the view of the Kelvin-Helmholtz instability discussed in \cite{choi2009regularized}, and the possibility of the exact solution derivation for the `regularized' two-fluid nonlinear one-dimensional model of \cite{choi2009regularized}. Another ongoing work direction is a systematic derivation of local conservation laws of the CC equations, and the comparison of the conserved densities with those for the full Euler model. The possible correspondence between the conserved quantities was suggested by R.~Camassa as a possible explanation of the quality of agreement between the solutions of the full Euler and the approximate Choi-Camassa models. A further interesting question is the possibility of derivation, for the Choi-Camassa model, of multi-soliton solutions similar to those known for various nonlinear equations of mathematical physics.

\subsubsection*{Acknowledgements}

The author is grateful to NSERC of Canada for the financial support of research through a Discovery grant.


\newpage

\bibliography{twofluid_ref32sol}
\bibliographystyle{ieeetr}

\begin{appendix}


\section{Coefficient Formulas for Theorem \ref{th:sol:snSQ}}\label{app:Th:snSQ}

The formula \eqref{eq:th:sol:elliptic:Z:snSQ} provides a solution of the ODE \eqref{eq:ODE:Z2} with \eqref{eq:ODE:Z2Ai} when the ODE and solution parameters are expressed in terms of $B_1$, $B_2$, $k$, $S$ though the formulas
\beq\label{eq:th:snSQ:flas1}
\barr
A_2=A_4\big[ B_1(2B_2+1)+3B_2(B_2+1)\big] +\dfrac{A_4B_1}{k^2}(B_1+2B_2+1),\\[3ex]
A_3=-A_4(B_1+3B_2+1) -\dfrac{A_4B_1}{k^2},\\[3ex]
\alpha_0=-\alpha_1=\dfrac{A_4 B_2}{3k^{2}} (B_1+B_2)(B_1+B_2k^2), \\[3ex]
\gamma^2 =  \dfrac{A_4 B_1}{4k^2\alpha_1},
\earr
\eeq
or through the formulas
\beq\label{eq:th:snSQ:flas2}
\barr
A_2=A_4B_2(2B_1+3B_2) +\dfrac{A_4B_1}{k^2}(B_1+2B_2),\\[3ex]
A_3=-A_4(B_1+3B_2) -\dfrac{A_4B_1}{k^2},\\[3ex]

\alpha_0=0,\qquad \alpha_1= -\dfrac{A_4}{3k^2}(B_2-1)(B_1+B_2-1)\big(B_1+k^2(B_2-1)\big),\\[3ex]

\gamma^2 = \dfrac{A_4 B_1}{4k^2\alpha_1}.
\earr
\eeq
In \eqref{eq:th:snSQ:flas1}, \eqref{eq:th:snSQ:flas2}, the constants $A_4$ and $A_1$ are determined by \eqref{eq:ODE:Z2Ai} and \eqref{eq:ODE:Z2Ai:A1}, respectively.

We note that for the solution form \eqref{eq:th:sol:elliptic:Z:snSQ}, the corresponding subfamilies of the ODEs \eqref{eq:ODE:Z2} are given by
\[
({\hat{Z}}')^2=\dfrac{ 3(B_2-\hat{Z})(B_1+B_2-\hat{Z}) (k^2(B_1+B_2-\hat{Z})-B_1)}{B_2 (B_1+B_2) (k^2(B_1+B_2)-B_1)},
\]
\[
({\hat{Z}}')^2=\dfrac{ 3(B_2-\hat{Z})(B_1+B_2-\hat{Z}) (k^2(B_1+B_2-\hat{Z})-B_1)}{(B_2-1) (B_1+B_2-1) (k^2(B_1+B_2-1)-B_1)},
\]
for the sets of constants \eqref{eq:th:snSQ:flas1} and \eqref{eq:th:snSQ:flas2}, respectively. 


\section{Coefficient Formulas for Theorem \ref{th:sol:sn:denom}}\label{app:Th:Sn:Den} 

The formula \eqref{eq:th:sol:elliptic:sn:denom} yields a solution of the ODE \eqref{eq:ODE:Z2} when the solution and the equation parameters $\gamma$, $k$, $\alpha_{0,1}$ are given in terms of the arbitrary constants $B_1, B_2, S$ by
\beq\label{eq:th:Den:flas1}
\barr
\alpha_0=-\alpha_1= -\dfrac{A_4 B_1^3}{6B_2(1-B_2^2)},\\[3ex]
\gamma^2 = \dfrac{3B_2^2}{B_1^2},\qquad  k^2=\dfrac{(1-(B_1-B_2)^2)}{B_2(2B_1-B_2)(B_1^2+(B_1-B_2)^2) + (B_1-B_2)^2},
\earr
\eeq
or through the formulas
\beq\label{eq:th:Den:flas2}
\barr
\alpha_0=0,\qquad \alpha_1=\dfrac{A_4 (2B_2-B_1)(1-(B_1-B_2)^2)}{6B_2(1-B_2^2)}, \\[3ex]
\gamma^2 =  \dfrac{3B_1B_2^2}{(2B_2-B_1)(1-(B_1-B_2)^2)},\qquad k^2=B_2^{-2},
\earr
\eeq
or through the formulas
\beq\label{eq:th:Den:flas3}
\barr
\alpha_0=\dfrac{A_4 B_1^3}{3(1-B_2^2)}\, \dfrac{1-(B_1-B_2)^2}{B_2(4B_1^2-5B_1B_2+2B_2^2) - 2B_2+B_1},\qquad \alpha_1=0, \\[3ex]
\gamma^2 =  \dfrac{3}{B_1^2}\,\dfrac{B_2(2B_1-B_2)(B_1^2+ (B_1-B_2)^2)+(B_1-B_2)^2}{1-(B_1-B_2)^2},\qquad k^2=\gamma^{-2}.
\earr
\eeq

In the expressions above, $A_4$ and $A_1$ are respectively determined by \eqref{eq:ODE:Z2Ai} and \eqref{eq:ODE:Z2Ai:A1}. Expressions for the constants $A_2$ and $A_3$ are straightforward to obtain; they are not listed here due to their complicated form and the lack of utility for writing down the physical solution components of the CC equations \eqref{eq:CC}.

\end{appendix}

\end{document}